\documentstyle[psfig,12pt,fleqn]{article} 
 
\begin{document} 
\title{ 
Integral methods for shallow free-surface\\ 
flows with separation 
} 
 
\author{ 
Shinya Watanabe\\ 
{\small Dept. of Mathematical Sciences, 
Ibaraki University, 310-8512, Mito, Japan}\\ 
Vachtang Putkaradze\\ 
{\small Dept. of Mathematics \& Statistics, 
University of New Mexico, 
Albuquerque, NM 87131-1141, USA}\\ 
Tomas Bohr\\ 
{\small Dept. of Physics, 
The Technical University of Denmark, Kgs. Lyngby, DK-2800, Denmark} 
} 
 
\date{ 
\normalsize Submitted on August 16, 2000.
}
 
\maketitle 
 
\begin{abstract} 
We study laminar thin film flows  
with large distortions in the free surface 
using the method of averaging across the flow. 
Two concrete problems are studied: 
the circular hydraulic jump and the flow down an inclined plane. 
For the circular hydraulic jump 
our method is able to handle an internal eddy 
and separated flow. 
Assuming a variable radial velocity profile 
like in Karman-Pohlhausen's method, 
we obtain a system of two ordinary differential equations 
for stationary states that can smoothly go through the jump 
where previous studies encountered a singularity. 
Solutions of the system are in good agreement 
with experiments. 
For the flow down an inclined plane we take a similar approach 
and derive a simple model in which the velocity profile is not 
restricted to a parabolic or self-similar form. 
Two types of solutions with large surface distortions are found: 
solitary, kink-like propagating fronts, obtained when the flow rate 
is suddenly changed, and 
stationary jumps, obtained, e.g., behind a sluice gate. 
We then include time-dependence in the model to study 
stability of these waves. 
This allows us to distinguish between sub- and supercritical flows 
by calculating dispersion relations  
for wavelengths of the order of the width of the layer.  
\end{abstract} 
 
%%%%%%%%%%%%%%%%%%%%%%% Introduction %%%%%%%%%%%%%%%%%%%%%%%%%%%%%%%%

\clearpage 
\section{Introduction} 
 
In this paper we develop a simple quantitative method 
to describe flows with a free surface which can undergo large 
distortions. 
Our method is capable of handling flows 
whose velocity profile may become far from parabolic --- even 
including separation and regions of reverse flow. 
We are concerned with the case 
when the fluid layer is thin. 
For low Reynolds number flows 
the lubrication approximation 
can be used with great success (see e.g. \cite{Eggers}). 
For high Reynolds number flows without separation 
an inviscid approximation and the shallow water equations 
\cite{Whitham} 
are widely used. 
For moderate Reynolds numbers 
where these limiting approximations are invalid 
it is important to take both inertial and viscous effects into account 
in a consistent way, and yet 
one would like to keep the model simple enough to be tractable. 
In this paper we show that integral methods, 
like the ones developed by von Karman, 
can handle a class of such problems successfully. 
To be concrete 
we develop the method in the context of two physical examples:  
the {\em circular hydraulic jump} and  
the {\em flow down an inclined plane}. 
Both geometries support jump- or kink-like solutions 
with abrupt changes in the surface shape 
and internal velocity profiles. 
Analytical solutions for such flows are extremely difficult to 
obtain, and 
simple approximate theories that capture the phenomena  
are invaluable. 

The two flows are studied in separate sections, 
and an introduction is provided in the beginning 
of Secs.~2 and 3, respectively. 
In Sec.~2 we develop the theory for the circular hydraulic jump. 
We first study the boundary layer approximation 
to the full Navier-Stokes equations, 
and reduce it to a simple set of equations 
by averaging over the thickness. 
Stationary solutions are obtained 
by solving a two-point boundary value problem 
for a system of only two ordinary differential equations. 
The solution is compared to previous experiments, 
showing  good agreement. 
Taking advantage of the simplicity of the reduced equations, 
it is possible to obtain analytic approximations 
for the stationary solution. 
Two ``outer'' solutions connected by an ``inner'' 
transition region are studied separately and 
we obtain a relationship analogous to the shock 
condition in the classical shock theory, 
but within our viscous model. 
 
The flow down an inclined plane is then studied 
in Sec.~3. 
We use the same strategy as in Sec.~2 
to derive a simple model for the two-dimensional flow. 
One family of solutions found in this model is 
kink-like traveling wave solutions that occur, e.g., 
when the flow rate is suddenly changed. 
Their velocity profiles along the inclined plane 
are found to stay close to parabolic 
even when a variable profile is assumed. 
There is another family of solutions 
with a sudden change in the surface 
that would correspond to the circular hydraulic jump 
in case of the radial geometry. 
These solutions can be interpreted as the stationary hydraulic jump, 
created behind a sluice gate in a river, 
even though turbulence is not included in the model. 
The flow downstream of the jump approaches a simple stationary 
flow, 
but the flow upstream is an expanding flow  
with a linear growth in thickness. 
The velocity profile departs considerably from parabolic 
near the jump. 
 
It is not easy to analyze the stability 
of the solutions with jumps obtained in Secs.~2 and 3, 
even in the linear geometry. 
Instead, in Sec.~4, we include time-dependence in the models 
and study the dispersion relation 
for the stationary flow with constant thickness. 
A well-established concept in the inviscid theory 
is to classify flows as super- and subcritical 
when the thickness is small and large, respectively. 
They do not have obvious counterparts, however, 
when viscosity is included. 
By looking carefully at the dispersion relation 
in the long and medium wave regime, 
we can classify the stationary flows into these two categories
in our viscous model.  The model shows spurious divergencies 
in the short wavelength region which we do not know how to 
overcome at present. This makes the model unsuited for direct 
time-dependent simulations. 
A short paper describing some of the main results has appeared 
earlier \cite{prelim}. 
 
%%%%%%%%%%%%%%%%%%%%%%%% Circular jump %%%%%%%%%%%%%%%%%%%%%%%%%%%%%%%% 
\clearpage 
\section{The circular hydraulic jump} 
 
\subsection{Introduction to the problem}  
 
\begin{figure} 
\centerline{
\psfig{file=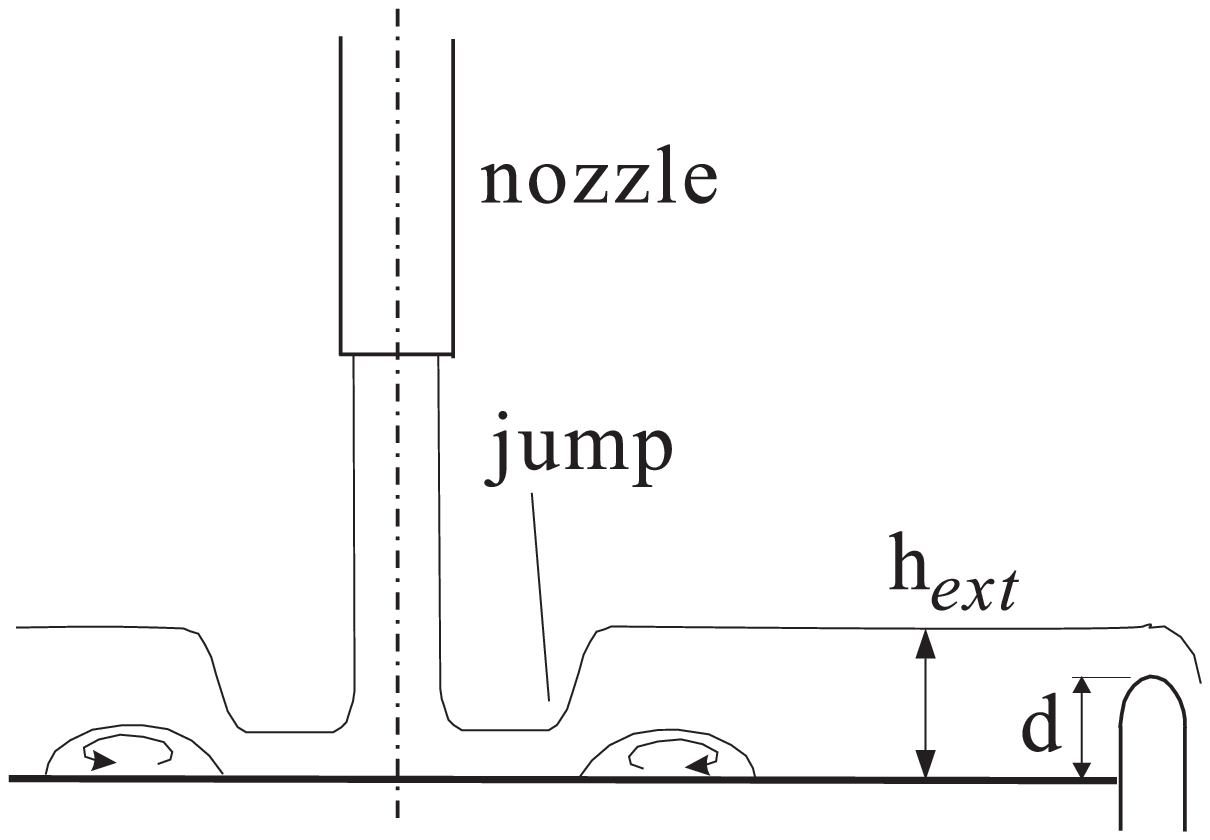,width=6cm} 
\psfig{file=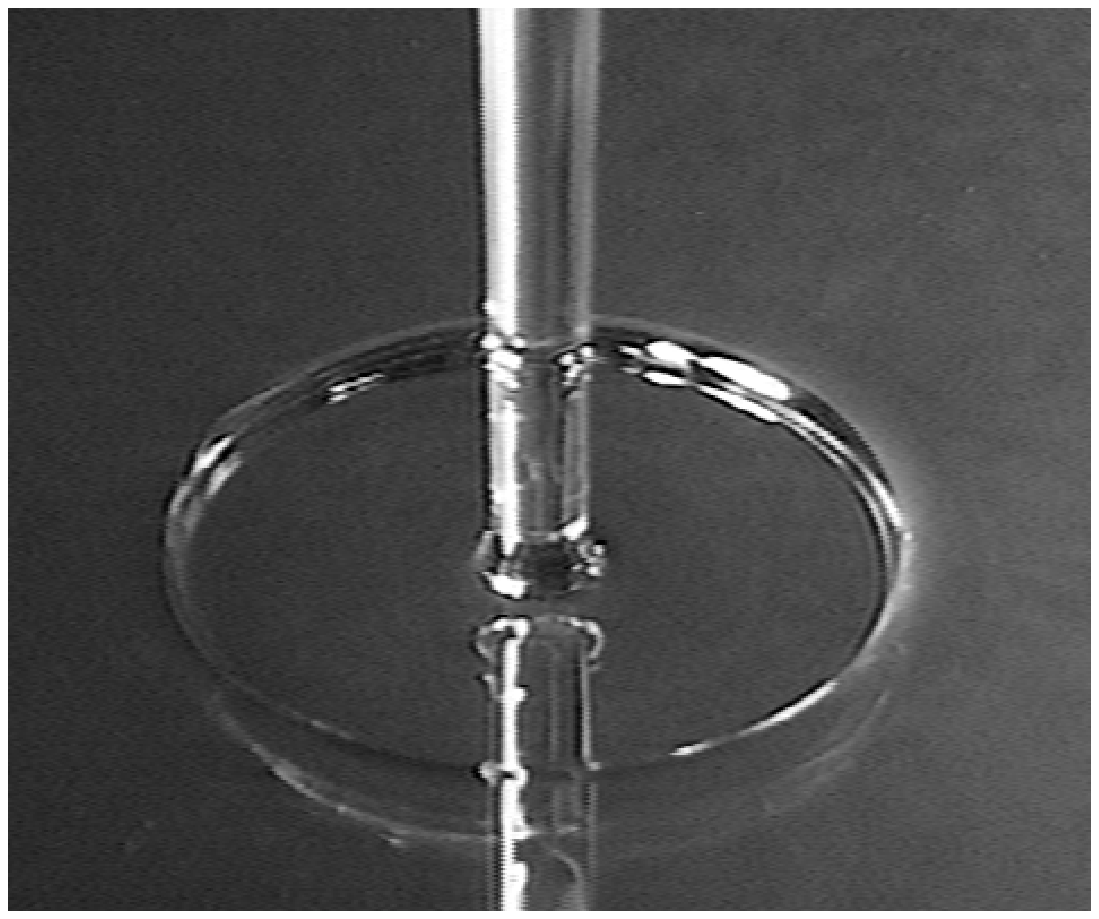,width=6cm}} 
\caption{ 
(a) 
Schematic view of the circular hydraulic jump. 
(b) 
Snapshot of a nearly perfect, stationary and circular 
hydraulic jump. 
Ethylene-glycol is used. 
} 
\label{expsetup} 
\end{figure} 
 
When a jet of fluid hits a flat horizontal surface, 
the fluid spreads out radially in a thin,  
rapidly flowing layer.  
At a certain distance from the jet a sudden thickening  
of the flow takes place, which is called the circular hydraulic jump. 
This is commonly seen, e.g., in the kitchen sink, 
but it is also important as a coating flow 
and in jet-cooling of a heated surface \cite{lienhard}. 
In these practical flows with typically high Reynolds numbers, 
disturbances often make the jump non-stationary and distorted. 
In controlled laboratory experiments 
corresponding to a more moderate Reynolds number, 
an apparently stationary, radially-symmetric flow can be achieved. 
Such experiments was carried out by C.\ Ellegaard {\em et al.} 
and the results have been published 
elsewhere \cite{behh,7samurai,polynature,polynonlin,Marcus}. 
We thank them for providing us with  data and pictures. 
A schematic view and a video image 
of the circular jump are shown in Fig.~\ref{expsetup}. 
 
\begin{figure}[t] 
  \centerline{\psfig{file=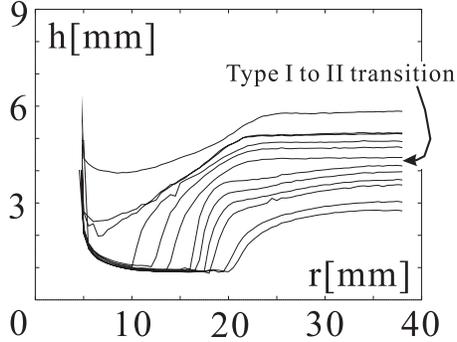,width=6cm}} 
\caption{ 
Height profiles $h(r)$ for different values 
of the external height $h_{ext}$. 
(The rim height $d$ is controlled but not shown.) 
The height $h(r)$ approaches $h_{ext}$ for large values of $r$. 
Parameters are:  
the flow rate $Q=27$[m$\ell$/s] and 
viscosity $\nu=7.6 \times 10^{-6}$[m$^2$/s], 
corresponding to the characteristic scales: 
radius $r_* = 2.8$[cm],  
height $h_* = 1.4$[mm], 
and radial velocity $u_* = 12$[cm/s]. 
Figure taken from \protect{\cite{behh}}. 
}
\label{series4} 
\end{figure} 
 
\begin{figure}[t] 
  \centerline{\psfig{file=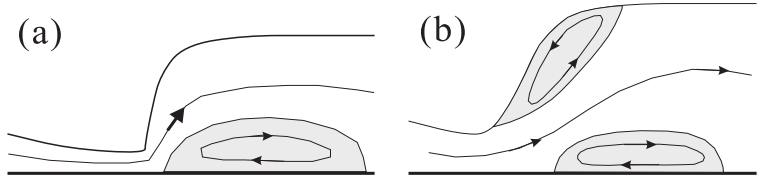,width=10cm}} 
\caption{ 
A schematic picture showing two observed flow patterns: 
(a) type I flow, with a separation bubble, which occurs for small 
$d$, and 
(b) type II flow, with an additional roller eddy, for large $d$. 
Transitions between these states occur at a certain $d$, 
with surprisingly small  hysteresis. 
} 
\label{type1-2} 
\end{figure}

In these experiments 
the hydraulic jump is formed on a flat disc with a circular rim. 
The rim height $d$ can be varied, 
and is an important control parameter. 
Since the rim is located far from the impinging jet 
with the diameter of the disc around 36cm, 
it does not affect the jump except that it 
changes the height of the fluid layer $h_{ext}$ 
exterior to the jump. 
The jump still forms even when $d=0$, 
but a larger $d$ makes $h_{ext}$ larger 
and, therefore, the jump stronger. 
Typically, $h_{ext}$ exceeds $d$ by 1-2mm. 
The surface profiles for varying $d$ are shown 
in Fig.~\ref{series4}. 
An interesting transition in the flow structure  
has been observed \cite{behh,7samurai} 
as $d$ is varied. 
For $d=0$, it was noticed before 
\cite{Tani,Craik,Ishigai,Nakoryakov,Olsson}  
that the jump contains an eddy on the bottom, 
called a {\em separation bubble}, 
whose inner edge is located very close to the position 
of the abrupt change on the surface, 
as illustrated in Fig.~\ref{type1-2}(a).  
Such a hydraulic jump is referred to as a {\em type I} jump. 
While $d$ remains small, this jump is stable, 
but as $d$ is increased further, 
a {\em wave-breaking} transition occurs \cite{behh,7samurai} 
which results in a new state of the flow. 
In this {\em type II} state, 
the flow has an additional eddy,  
called a {\em roller} or a {\em surfing wave}, 
just under the surface as shown 
in Fig.~\ref{type1-2}(b).\footnote{ 
If $d$ is increased even further, 
the jump ``closes'' as seen in Fig.~\ref{series4}. 
} 
This state resembles a broken wave in the ocean, 
but is still apparently laminar. 
On reducing $d$, the type I pattern reappears, 
and there is almost no hysteresis associated with this transition. 
The transition from type I to II often leads also to 
breaking of the radial symmetry. 
An intriguing set of polygonal jumps \cite{polynature,polynonlin} 
are created rather than the circular one. 
In this paper we shall concentrate on the type I flow 
which already poses considerable difficulties. 
We hope to be able to generalize our approach in the future 
to be able to handle the transition to the type II flow. 
 
Considering how simple and common the circular hydraulic jump 
appears to be, 
it is surprising that a satisfactory systematic theory  
does not exist. 
The approach considered as ``the standard'' 
for the study of hydraulic jumps is 
to combine the inviscid shallow water equation 
with Rayleigh's shocks \cite{Chow}. 
In the beginning of the century Lord Rayleigh 
treated \cite{Rayleigh} a discontinuity 
in a one-dimensional linear flow geometry. 
Such a structure is usually called a {\em river bore} if it is moving  
and a {\em hydraulic jump} 
if it is stationary and is created due to, e.g., 
variations in the river bed. 
His approach was based upon the analogy  
between the shallow water theory and gas theory \cite{Whitham}. 
He assumed that, across such a shock, 
the mass and momentum flux are conserved 
but not the energy flux. 
 
In a coordinate system moving with the shock, 
the flow velocity $v_1$ and height $h_1$ upstream of the jump 
as well as $v_2$ and $h_2$ downstream of the jump are taken to be 
positive constant values. 
Then, conservation of mass flux $Q$ 
across the jump is given by 
\begin{equation} 
  \label{cflux} 
 v_1 h_1 = v_2 h_2 = Q . 
\end{equation} 
Conservation of momentum flux is 
\begin{equation} 
  \label{cmomentum} 
  h_1 \left( v_1^{2}+ \frac{1}{2} gh_1 \right)  
 = h_2 \left( v_2^{2}+ \frac{1}{2} gh_2 \right).   
\end{equation} 
These shock conditions lead to the relation 
\begin{equation} 
  \label{Froude} 
{\frac{h_2}{h_1}} = {\frac{1}{2}}(-1 + \sqrt{1 + 8 F_1^2}) 
  = \frac{2}{-1 + \sqrt{1 + 8 F_2^2}} 
\end{equation} 
where $F_1 = \sqrt{v_1^2/g h_1} = ( h_c/h_1 )^{3/2}$ is the 
{\em upstream Froude number}, 
$F_2 = \sqrt{v_2^2/g h_2} = ( h_c/h_2 )^{3/2}$ the {\em downstream 
Froude number}, 
and  
\begin{equation} 
  h_c = ( Q^2/g )^{1/3} 
\label{criticalheight} 
\end{equation} 
is called the critical height. 
It is easy to see that 
$h_c$ is always between $h_1$ and $h_2$, 
and that 
$F_1 > 1 > F_2$ if $h_1 < h_c < h_2$, 
and 
$F_1 < 1 < F_2$ if $h_1 > h_c > h_2$. 
In other words the jump connects 
a {\em supercritical flow} with $F>1$ on the shallower side 
($h<h_c$) 
to a {\em subcritical flow} with $F<1$ on the deeper side 
($h>h_c$). 
Since the Froude number measures 
the ratio of the fluid velocity $v$ and 
the velocity of linear surface waves $\sqrt{gh}$, 
it means that, in the moving frame, 
the flow moves more rapidly 
than the surface waves on the shallower side, 
but moves slower on the deeper side --- 
in a precise analogy with the gas theory \cite{Whitham,Stoker}.  
Further, it is found that the upstream $h_1$ must be supercritical 
by considering the change in the energy flux 
across the jump \cite{Whitham}: 
\begin{equation} 
  \label{dissenr} 
  Q_{e2} - Q_{e1} = - \frac{gQ(h_{2}-h_{1})^{3}} 
  {8 \pi h_{1} h_{2}} 
\end{equation} 
where $Q_{e}$ denotes the energy flux. 
Since the energy must be dissipated through the jump, 
i.e.\ $Q_{e2} - Q_{e1} < 0$, rather than generated, 
it is required that $h_1 < h_2$. 
The origin of the dissipation is usually attributed to  
the turbulent motions at the discontinuity 
and surface waves carrying energy away from it. 
 
It is possible to apply this theory, 
combined with an assumption of the potential flow, 
for describing the circular hydraulic jump. 
However, it leads to incorrect estimates \cite{Watson,BDP} 
of the radius of the jump $R_j$.  
Most notably, $R_j$ is predicted to be sensitive 
to the radius of the impinging jet 
which should be greatly influenced by radius and height 
of the inlet nozzle where liquid comes out. 
In experiments \cite{Watson,BDP} such a strong tendency 
was not observed. 
Instead, it has been found that $R_j$ scales 
with the flow rate $Q$ with a certain power, 
and it supports a model in which viscosity plays an important role. 
Watson \cite{Watson} constructed a model of the flow consisting of 
the inviscid and viscid regimes, 
and solved the viscid part assuming a similarity profile. 
By connecting to the specified external height $h_{ext}$ 
via a Rayleigh shock, 
he obtained a prediction for the radius of the jump 
which compares favorably with the measurement 
\cite{Watson,BDP}, 
as we explain in Sec.~2.5. 
In his model the viscous layer starts from the stagnation point 
at $r=0$ on the plate and quickly reaches the surface 
at a small $r$. 
There is a fairly long stretch from this $r$ to $R_j$ 
in which the flow is fully viscous.\footnote{ 
This assumption is confirmed by recent laser-doppler 
measurements 
of the velocity profile before the jump \cite{Marcus}. 
Thus, the 
assumption made by  \cite{Godwin,Blackford,Brechet} 
that the jump occurs at the point where the growing viscous layer 
touches the surface and the flow becomes fully developed, is 
incorrect. } 
Thus, one could neglect the inviscid region 
and assume a fully viscid flow everywhere 
in order to derive a simpler model. 
This assumption was made by Kurihara \cite{Kurihara} 
and Tani \cite{Tani} who started 
from the boundary layer equations developed by Prandtl 
\cite{Prandtl,Schlichting}. 
They took an average of the equations over the thickness, 
assuming also a similarity velocity profile. 
It resulted in a single ordinary differential equation 
for the stationary jump. 
This theory was elaborated in \cite{BDP} 
who realized that the flow outside the jump would naturally 
lead to a singularity at a large $r$. 
By identifying this singularity with the outflow 
over the rim of the plate, 
the flow outside the jump could be uniquely specified. 
By introducing a Rayleigh shock, 
the jump radius and its parameter dependence was calculated 
and compared to measurements. 
The model predicted the observed $R_j$ reasonably well, 
as we review in Secs.~2.2--2.5. 
 
Obviously, treating the jump as a discontinuity provides us with 
no information on the internal structure of the jump region 
such as the type I to II transition of the flow patterns. 
It also seems inconsistent to assume a Rayleigh shock 
when viscous loss occurs in the whole domain.
Why do we assume an extra energy loss at the ``jump'' 
where the flow is stationary and apparently laminar? 
It seems possible to attribute the energy dissipation 
entirely to laminar viscous forces, 
and to construct a viscous theory 
which produces a smooth but kink-like surface shape 
without the need for a discontinuity. 
Nevertheless, such a description must overcome a difficulty 
arising from the Goldstein-type singularity \cite{Goldstein,Landau} 
of the boundary layer equations in the vicinity of separation 
points. 
This singularity is thought to be an artificial one 
created by truncation of higher derivatives 
from the Navier-Stokes equations. 
It also arises in the ``usual'' boundary layer situation 
where a high Reynolds number flow passes a body, e.g., a wing. 
In such cases inviscid-viscid interaction is taken into account 
in order to resolve the singularity 
in a technique called the inverse method \cite{inverse}. 
In our situation, however,  
there is no inviscid flow outside the layer. 
In Secs.~2.6--2.7 we propose a way to resolve the trouble 
in the following manner. 
We first include an additional degree of freedom 
in the velocity profile to make it non-self-similar, 
just like in the Karman-Pohlhausen method \cite{integralmethod} 
for the usual boundary layer theory. 
To describe the evolution, in $r$, of this free parameter, 
we couple the layer thickness to the pressure 
by assuming  hydrostatic pressure. 
This serves as an alternative to the inverse method 
in the absence of  a potential external flow. 
The resulting model for a stationary solution 
is two coupled ordinary differential equations, 
and reproduces the type I flow with a separation bubble --- the one 
shown in Fig.~\ref{type1-2}(a). 
Comparison with the experiment is made in Sec.~2.7.
It is possible to approximate analytically the stationary solution 
found in the model. 
In Sec.~2.8 the analysis is presented separately for 
the regions before and after the jump 
(i.e.\ two ``outer'' solutions) 
and the ``inner'' solution inside the jump region. 
An interesting observation on the inner solution 
is that a formal parameter $\mu$ can be introduced 
so that Rayleigh's shock condition is recovered 
in the limit $\mu \rightarrow 0$. 
 
\subsection{The full model} 
 
We write down 
the complete model to describe the circular hydraulic jump 
under the assumption that the flow is laminar and 
radially symmetric 
without any angular velocity component. 
We take the radial and vertical coordinates $\tilde{r}$ and 
$\tilde{z}$, 
and denote the velocity components 
by $\tilde{u}$ and $\tilde{w}$, respectively.\footnote{ 
We use tildes  
for the dimensional variables, dependent or independent. 
Dimensionless variables will be expressed 
by the same symbols but without tildes. 
In figures, however, we do not use tildes for simplicity.} 
The governing equations are the continuity equation 
\begin{equation}  
  \tilde{u}_{\tilde{r}} + \frac{\tilde{u}}{\tilde{r}} +  
  \tilde{w}_{\tilde{z}} = 0 
\label{dimincompr}  
\end{equation}  
and the Navier-Stokes equations: 
\begin{equation} 
  \begin{array}{l}  
    \displaystyle 
    \tilde{u}_{\tilde{t}} +  
    \tilde{u} \tilde{u}_{\tilde{r}} +  
      \tilde{w} \tilde{u}_{\tilde{z}} =  
     -\frac{1}{\rho} \tilde{p}_{\tilde{r}} 
     +\nu \left( \tilde{u}_{\tilde{r} \tilde{r}} 
     +\frac{1}{\tilde{r}} \tilde{u}_{\tilde{r}} 
     -\frac{\tilde{u}}{\tilde{r}^2} 
     +\tilde{u}_{\tilde{z} \tilde{z}} \right) \\  
    \displaystyle 
    \tilde{w}_{\tilde{t}} +  
    \tilde{u} \tilde{w}_{\tilde{r}} 
     +\tilde{w} \tilde{w}_{\tilde{z}} =  
     -\frac{1}{\rho} \tilde{p}_{\tilde{z}} -g 
     +\nu \left( \tilde{w}_{\tilde{r} \tilde{r}} 
     +\frac{1}{\tilde{r}} \tilde{w}_{\tilde{r}} 
     +\tilde{w}_{\tilde{z} \tilde{z}} \right) 
  \end{array}  
\label{dimNS} 
\end{equation}  
where subscripts 
denote partial differentiations 
such as  
$\displaystyle \tilde{u}_{\tilde{t}} = 
\partial \tilde{u} / \partial \tilde{t}$. 
For the boundary conditions we impose no-slip on the bottom:  
\begin{equation}  
  \tilde{u}(\tilde{z}=0)=\tilde{w} (\tilde{z}=0)=0.  
\label{no-slipdim} 
\end{equation} 
The dynamic boundary conditions on the free surface 
$\tilde{z}=\tilde{h}(\tilde{t},\tilde{r})$ are  
\begin{equation}  
  \begin{array}{l}   
    \displaystyle 
    \tilde{p}-\frac{2 \rho \nu}{1+\tilde{h}_{\tilde{r}}^2}  
      \left. \left[ 
        \tilde{h}_{\tilde{r}}^2 \tilde{u}_{\tilde{r}}+  
        \tilde{w}_{\tilde{z}}-  
        2 \tilde{h}_{\tilde{r}} \left( 
          \tilde{w}_{\tilde{r}} + \tilde{u}_{\tilde{z}} 
        \right) 
      \right] \right|_{\tilde{z}=\tilde{h}} 
    =\sigma \tilde{k} \\  
    \displaystyle 
    \nu \left. \left[ 
      \left( \tilde{h}_{\tilde{r}}^2 -1 \right) \left( 
        \tilde{w}_{\tilde{r}} + \tilde{u}_{\tilde{z}} 
      \right) -  
      2 \tilde{h}_{\tilde{r}} \left( 
        \tilde{u}_{\tilde{r}} - \tilde{w}_{\tilde{z}} 
      \right) 
    \right] \right|_{\tilde{z}=\tilde{h}} 
    = 0  
  \end{array}  
\label{dimBC} 
\end{equation}  
where $\sigma$ is the coefficient of surface tension and $\tilde{k}$ 
is the  
mean local curvature of the free surface. 
We also need to satisfy the kinematic boundary condition 
on the free surface: 
\begin{equation} 
  \tilde{h}_{\tilde{t}} + \tilde{u} \tilde{h}_{\tilde{r}} = \tilde{w} 
  \qquad \mbox{on $\tilde{z} = \tilde{h} (\tilde{t},\tilde{r})$.} 
\label{kinematicbcdim} 
\end{equation} 
We are mostly interested in stationary solutions in this section. 
When the flow is stationary, 
we may integrate (\ref{dimincompr})  
over $\tilde{z}$ from 0 to $\tilde{h}$, 
and use (\ref{kinematicbcdim}) to obtain 
\begin{equation} 
  \tilde{r} \displaystyle \int^{\tilde{h}(\tilde{r})}_{0} 
  \tilde{u}(\tilde{r} , \tilde{z}) \mbox{d}\tilde{z} 
  = q = \frac{Q}{2 \pi} . 
\label{nonrescmass} 
\end{equation} 
This quantity, the total mass flux $Q$ 
or the mass flux per angle $q$, 
is a constant, given as a parameter in the experiment. 
 
\subsection{Boundary layer approximation} 
 
Since it is a formidable task to treat the full model as it stands, 
some simplifications need to be made. 
As explained in Sec.~1, 
the Reynolds number for the flow of the circular hydraulic jump 
is too large to justify the lubrication approximation, 
but is not large enough to use the inviscid approximation. 
Fortunately, the flow is ``thin,'' i.e.\ runs 
predominantly horizontally along the plate. 
Truncation of the full model by the boundary layer approximation 
is quite natural in such a situation, 
and has indeed been used in previous 
literature \cite{Kurihara,Tani,BDP}. 
In the boundary layer approximation 
pressure, viscous, and inertial terms in (\ref{dimNS}) 
are all assumed to be of the same order, 
but there are only a few dominant terms in each group. 
For instance, a viscous term  
$\nu \tilde{u}_{\tilde{r} \tilde{r}}$ 
is assumed to be negligible compared to 
$\nu \tilde{u}_{\tilde{z} \tilde{z}}$. 
The dominant terms in the first equation in (\ref{dimNS}) 
are determined in the usual manner: 
$\tilde{u}_{\tilde{t}}$ (if time-dependent), 
inertia terms  
$\tilde{u} \tilde{u}_{\tilde{r}}$ and 
$\tilde{w} \tilde{u}_{\tilde{z}}$, 
the pressure term 
$\tilde{p}_{\tilde{r}}/\rho$, 
and the dominant viscous term 
$\nu \tilde{u}_{\tilde{z} \tilde{z}}$. 
Similarly, from the second equation in (\ref{dimNS}) 
we assume the dominant balance between 
$\tilde{p}_{\tilde{z}}/\rho$ and $g$. 
Here, unlike the usual boundary layer theory, 
we have taken into account the effect of gravity. 
This will couple the surface height $h$ to the pressure, 
and will later turn out to be crucial for removing the singularities 
of the boundary layer approximation. 
 
If we denote the characteristic radius and height 
by $r_*$ and $z_*$, respectively, 
then the second dominant balance requires 
the characteristic pressure to be $\rho g z_*$. 
Then, the first balance relation requires 
\begin{equation} 
  \frac{u_*}{t_*} = \frac{u_*^2}{r_*} = \frac{u_* w_*}{z_*} 
  = \frac{\rho g z_*}{\rho r_*} = \frac{\nu u_*}{z_*^2}  
\label{dominantbalance} 
\end{equation} 
where $u_*$ and $w_*$ are typical radial and vertical velocities, 
respectively, and $t_*$ is the characteristic time scale. 
The mass flux relation (\ref{nonrescmass}) requires that  
\begin{equation}  
  u_* r_* z_*=q.  
\label{massconsscale} 
\end{equation}   
while the continuity equation (\ref{dimincompr}) requires  
\begin{equation}  
  \frac{u_*}{r_*}=\frac{w_*}{z_*} . 
\label{incomprscaling} 
\end{equation} 
Solving (\ref{dominantbalance}),  
(\ref{massconsscale}), and (\ref{incomprscaling}) 
uniquely determines the characteristic scales: 
\begin{equation} 
  \begin{array}{l}  
    r_*=\left( q^5  \nu^{-3} g^{-1} \right)^{1/8} \simeq 2.7 
\mbox{[cm]} , \\ 
    z_*=\left( q \nu g^{-1} \right)^{1/4} \simeq 1.5 \mbox{[mm]} , 
\\ 
    u_*=\left( q \nu g^3 \right)^{1/8} \simeq 12 \mbox{[cm/s]} , \\ 
    w_*=\left( q^{-1} \nu^3 g \right)^{1/4} \simeq 6.7 
\mbox{[mm/s]} , \\ 
    t_*=\left( q \nu^{-1} g^{-1} \right)^{1/2} \simeq 0.22 
\mbox{[s]} 
\end{array} 
\label{nondimrescaling} 
\end{equation} 
where the estimated values correspond to a typical set of  
parameters used in the experiments: 
$\nu \simeq 0.1$ cm$^2$/s (for mixture of ethylene-glycol and 
water) 
and $Q \simeq 30$ cm$^3$/s, i.e., $q \simeq 5$ cm$^3$/s. 
The values for $r_*$ and $z_*$ correspond well to 
a typical jump radius and fluid thickness in the experiments. 
Also, the predicted scaling can be experimentally tested by, 
for instance, measuring the dependence of the jump radius 
by changing the parameters such as $q$. 
In \cite{BDP} evidence of the scaling 
and validity of the underlying assumption was given. 
 
We now use the characteristic scales (\ref{nondimrescaling}),
together with the pressure scale $p_*=\rho u_*^2$,
to non-dimensionalize the full equations. 
From (\ref{dimNS}), we obtain 
\begin{equation}  
  \begin{array}{l} 
    \displaystyle 
    u_t + u u_r + w u_z = -p_r + u_{zz} + \epsilon^2 \left( 
      u_{rr} + \frac{1}{r} u_r - \frac{u}{r^2} 
    \right) \\  
    \displaystyle 
    \epsilon^2 \left( w_t + u w_r + w w_z \right) =  
      -p_z - 1 + \epsilon^2 w_{zz} + 
      \epsilon^4 \left( w_{rr} + \frac{1}{r} w_r \right) ,   
  \end{array}  
\label{nondimNS} 
\end{equation}  
where  
\begin{equation} 
  \epsilon=z_*/r_*=\left( q^{-3} \nu^5 g^{-1} \right)^{1/8} . 
\label{eps}  
\end{equation} 
Since $\epsilon = z_*/r_* = 0.05$ 
for the typical parameter values above, 
the assumption that the flow is ``thin'' is well satisfied, 
and we shall drop the terms of order $\epsilon^2$ and higher 
in the equations (\ref{nondimNS}). 
We also focus on stationary solutions in the rest of the section, 
and thus we obtain the simplified equations of motion: 
\begin{equation}  
  \begin{array}{l}   
    u u_r + w u_z = - p_r + u_{zz} \\ 
    0 = -p_z - 1 . 
  \end{array}  
\label{nondimBLeeq0} 
\end{equation}  
Correspondingly, within the error of $O(\epsilon^2)$, 
the dynamic boundary conditions (\ref{dimBC}) are just  
\begin{equation} 
  \begin{array}{l}   
  \left. p \right|_{z=h} = W h_{rr} \\ 
  \left. u_z \right|_{z=h} = 0 . 
  \end{array}  
\label{dynamicbc}  
\end{equation} 
Here we have introduced the Weber number 
\begin{equation}  
  W = \frac{\sigma z_*}{\rho u_*^2 r_*^2} 
  = \frac{\sigma }{\rho g r_*^2} 
  = \frac{\ell^2}{2 r_*^2} 
  = \sigma \rho^{-1} ( q^{-5} \nu^{3} g^{-3} )^{1/4} . 
\label{webernumber} 
\end{equation}  
where $\ell = ( 2 \sigma/ (g \rho ))^{1/2}$ is the capillary length. 
For the parameter values above 
together with $\sigma \sim 70$[dyn/cm] (maximum), 
we estimate that $W \sim 0.01$ 
and $\ell \sim 3.8$[mm]. 
Since $W$ is small, we neglect it in the study of stationary 
states.\footnote{ 
However, the term influences dispersion of short waves, 
so should be included in the stability analysis of stationary states, 
possibly together with the neglected terms of $O(\epsilon^2)$ and 
higher 
in (\ref{nondimBLeeq0}).}$^,$\footnote{ 
The Reynolds number, defined as 
$R = u_* z_*/\nu = (q^3 \nu^{-5} g)^{1/8} \approx 18$. 
The Reynolds number at the nozzle outlet is much higher, 
but it becomes moderate near the jump.} 
The second equation of (\ref{nondimBLeeq0}) and the first  
condition of (\ref{dynamicbc}) with $W$ set to zero 
yield hydrostatic pressure:  
\begin{equation} 
  p(r,z) = h(r)-z . 
\label{hydrostaticpressure} 
\end{equation} 
Combining (\ref{nondimBLeeq0}) and (\ref{hydrostaticpressure}),   
we obtain the stationary boundary layer equations: 
\begin{equation} 
  u u_r + w u_z = -h' + u_{zz} , 
\label{rescBLrad} 
\end{equation} 
where the prime denotes the derivative with respect to $r$. 
This is supplemented by the dimensionless continuity equation: 
\begin{equation}  
  u_r + \frac{u}{r} + w_z = 0 , 
\label{incompressibility}  
\end{equation}  
and mass flux condition: 
\begin{equation}  
  r \int_0^{h(r)} u(r,z) dz =1 . 
\label{dimflux}  
\end{equation}  
The boundary conditions have been reduced to: 
\begin{equation} 
  \begin{array}{l} 
    u(r,0)=w(r,0)=0  \\ 
    \displaystyle \left. u_z \right|_{z=h(r)} = 0 . 
\end{array} 
\label{rescbc} 
\end{equation}   
In addition to these conditions, 
boundary conditions in the radial direction also need to be 
specified. 
We do not elaborate on them, however, 
since the in- and outlet conditions arise naturally 
without the need for prescription 
when we obtain a simplified system. 
 
The boundary layer equations (\ref{rescBLrad})--(\ref{rescbc}) 
form a closed system and can be solved numerically, 
but pose a difficulty when separated regions exist. 
Suppose that there is a separation point at $r=r_s$ and $z=0$ 
on a flat plate 
where the skin friction $u_z$ vanishes. 
In its vicinity one finds \cite{Schlichting} that 
generic solutions of (\ref{rescBLrad}) develop singularities 
of the Goldstein-type 
$u \sim \sqrt{r_s-r}$, $w \sim 1/\sqrt{r_s-r}$. 
On the other hand, experiments \cite{behh} show 
separation and reversed flow just behind a jump, 
so it is necessary to overcome this difficulty, 
which is well-known in the ``usual'' boundary layers 
around a body immersed in a high Reynolds number external flow. 
No such singularities are observed in numerics  
of the full Navier-Stokes equations in that case, 
and thus the trouble is thought to be due to truncation 
of the terms involving higher derivatives in $r$, 
i.e.\ the terms in (\ref{nondimNS})  
of the order $O(\epsilon^2)$ and higher. 
An attempt to include those terms leads to intractable equations, 
so the inverse method \cite{inverse} is often used. 
In this method the feedback from the boundary layer 
into the external potential flow is taken into account, 
and the coupled system is iteratively solved to remove the 
singularity. 
Without such an external flow present for the circular hydraulic 
jump, 
Higuera \cite{Higuera} has still obtained the velocity and height 
profiles  
from the boundary layer equations. 
His method, called marginal separation, 
is to force the boundary layer equations 
through the point of separation by choosing  
a special non-divergent velocity profile at the point. 
The physical reasoning for the choice of such a particular profile 
is rather unclear. 
Since our aim is also to obtain a simple tractable model, 
we have chosen a different approach. 
 
\subsection{Averaged equations} 
 
Rather than solving the partial differential equation 
(\ref{rescBLrad}) itself, 
we shall be content with satisfying only the 
mass and momentum conservation laws, 
derived from averaging (\ref{rescBLrad}) 
over the transverse $z$-direction. 
To do this we make an ansatz for the radial velocity profile $u$. 
One might expect that 
the singularities at separation points do not contribute 
to the averages and do not cause any harm. 
Such an expectation is too naive as shown in the next section, 
since the model still shows singular behavior 
near the jump if the simplest velocity profile is assumed. 
Nevertheless, we show in Sec.~2.7 that 
the model becomes capable of going through the jump smoothly 
once enough flexibility is introduced in the assumed profile. 
 
We first define the average velocity at $r$ by 
\begin{equation}  
  v=\frac{1}{h} \int^h_0 u(r,z) dz . 
\label{avervel}  
\end{equation} 
The total mass flux condition (\ref{dimflux}) can be written as 
\begin{equation}  
  rhv =1 . 
\label{avvelflux}  
\end{equation}   
Next, for each fixed $r$, 
we integrate the radial momentum equation (\ref{rescBLrad}) 
over $z$ from 0 to $h(r)$, 
and use the continuity equation (\ref{incompressibility})  
with the surface boundary conditions (\ref{rescbc}). 
We obtain the averaged momentum equation 
\begin{equation} 
  \frac{1}{rh} \frac{d}{dr} \left[ r \int_0^h u^2 \mbox{d}z \right] 
= 
  - h' - \frac{1}{h} \left. u_z \right|_{z=0} . 
\label{intmomeq}  
\end{equation} 
Using $v$ and  
\begin{equation} 
  G = \frac{1}{h} \int_0^h \left( \frac{u}{v} \right)^2 dz , 
\label{squareaverage} 
\end{equation} 
we obtain 
\begin{equation} 
  v (G v)' = 
  - h' - \frac{1}{h} \left. u_z \right|_{z=0} . 
\label{averagedmomentum} 
\end{equation} 
Equations (\ref{avvelflux}) and (\ref{averagedmomentum}) 
are the total mass and momentum equations. 
 
\subsection{Similarity profile for $u$} 
\label{sec:similarity} 
 
The simplest assumption for the radial velocity profile is 
a self-similar ansatz: 
\begin{equation} 
  u(r,z)/v(r) = f(\eta) 
\label{similarityprofile} 
\end{equation} 
where $\eta = z/h(r)$ takes values between 0 (bottom) 
and 1 (surface). 
Using (\ref{incompressibility}),  
the ansatz can be rewritten in the alternate form:  
$w(r,z)= \eta h' u(r,z)$.  
It is also equivalent to the requirement that 
the local inclination of the streamlines at $(r,z)$  
be proportional to $\eta h'=z h'(r)/h(r)$.  
Clearly, such an ansatz is too simple and ``rigid'' 
to describe a flow with separation.  
However, this is the assumption used in the previous literature, 
and we summarize its consequences. 
For more details, see \cite{BDP}.  
  
The conditions (\ref{rescbc}) and (\ref{avvelflux}) 
now imply 
\begin{equation} 
\begin{array}{l}  
  f(0)=0 , \\  
  f'(1)=0 , \\ 
  \displaystyle \int_0^1 f(\eta) d\eta = 1 . 
\end{array}  
\label{profileconstraints} 
\end{equation} 
They are not sufficient to uniquely determine $f$, 
and we choose one that is physically reasonable. 
Thus, a parabolic profile $ f(\eta) = 3 \eta - 3/2 \eta^2$ 
is a simple candidate. 
Using this choice, 
$G = 6/5$ is a constant from (\ref{squareaverage}), 
and (\ref{averagedmomentum}) becomes 
\begin{equation} 
  \frac{6}{5}v  v' = -h' - \frac{3 v}{h^2} . 
\label{Tanieqn} 
\end{equation} 
Other choices for $f$ lead to the same equation 
with different numerical coefficients. 
Since all such equations, 
corresponding to different choices of $f(\eta)$, 
can be further transformed to 
\begin{equation} 
  v v' = -h' - \frac{v}{h^2} 
\label{scaledTanieqn} 
\end{equation} 
by suitably including numerical coefficients 
in the characteristic scales (\ref{nondimrescaling}), 
the choice of $f$ is not important 
in the study of qualitative behaviour 
and of parameter dependence. 
 
Using (\ref{avvelflux}), 
the equation reduces to a single ordinary differential equation  
for $v(r)$: 
\begin{equation} 
  v' \left(v-\frac{1}{v^2 r} \right) = \frac{1}{v r^2}  - v^3 r^2 . 
\label{tanieqv} 
\end{equation} 
This Kurihara-Tani equation was derived and studied  
in \cite{Tani}, in its dimensional form, and in \cite{BDP}. 
The results can be summarized as follows. 
To find a solution corresponding to a hydraulic jump, 
the velocity $v$ should be large for small $r$, 
and decrease smoothly as $r$ increases. 
However, the model does not have such a solution. 
The coefficient of $v'$ on the left hand side generically vanishes 
at some $r$ where $v'$ diverges. 
If (\ref{tanieqv}) is solved in a parametric form 
on the $(r,v)$-plane, 
all solutions spiral around and into the fixed point $(r,v)=(1,1)$, 
that is a stable focus in the plane. 
Therefore, one must still connect solutions 
in the interior and the exterior by means of, e.g., 
a Rayleigh shock across which mass  
and momentum flux are conserved. 
When this is carried out, 
one finds that the shock occurs very close to $r=1$ 
in the dimensionless coordinates, 
implying that the radius of the jump in the dimensional coordinates 
scales roughly as $r_*$ in (\ref{nondimrescaling}), i.e.: 
\begin{equation}  
  R_{j} \propto \left( q^5 \nu^{-3} g^{-1} \right)^{1/8} .   
\label{jumpscale}  
\end{equation} 
This scaling relation (\ref{jumpscale}) was compared to 
experiments \cite{BDP,Dimon} by changing $q$ for several different 
$\nu$. 
The radius of the jump indeed scaled with the mass flux $q$, 
but the exponent observed in the experiment was about 3/4 
rather than 5/8 suggested by (\ref{jumpscale}). 
To explain the discrepancy, $R_j$ was calculated 
more accurately \cite{BDP}. 
It was first proven that there is no solution for $v(r)$ 
to the Kurihara-Tani equation that extends to $r=\infty$. 
All solutions were found to diverge 
at some $r=r_{\mbox{\scriptsize end}}$ (constant) 
like $h \sim \left\{ \log (r_{\mbox{\scriptsize end}}/r) 
\right\}^{1/4}$. 
By identifying this singularity as the end of the plate 
where the water runs off, 
one may always find the solution of (\ref{tanieqv}) 
diverging at the end of the plate of a given radius 
$r=r_{\mbox{\scriptsize end}}$. 
By following the solution to smaller $r$, 
the solution before the jump and the  
position of the shock are uniquely determined 
assuming a connection via a Rayleigh shock. 
The shock location constructed in this way showed 
a good agreement \cite{BDP,Dimon} with the experiment. 
 
\subsection{Profile with a shape parameter} 
 
An ansatz more flexible than (\ref{similarityprofile}) must be used  
for resolving the flow pattern in the vicinity of the jump. 
We shall allow the function $f$ in (\ref{similarityprofile}) 
to depend also on $r$. 
The simplest modification we can make is to 
assume $f = f(\eta,\lambda(r))$ 
so that the velocity profile is characterized 
by a single ``shape parameter'' $\lambda(r)$. 
The approach follows the ideas developed 
by von Karman and Pohlhausen \cite{Schlichting} 
for the usual boundary layer flow around a body. 
There, separation of the boundary layer can occur 
when the pressure gradient, 
imposed by the external inviscid flow, 
becomes adverse. 
In our case, there is no external flow, 
but there is a pressure gradient, along the bottom $z=0$, 
that is proportional to $h'(r)$ 
due to the hydrostatic pressure (\ref{hydrostaticpressure}).  
Thus, the possibility arises that the flow separates on $z=0$ 
near the jump where $h'$ is large and pressure is  
increasing in $r$, 
as in the usual boundary layer flow. 
 
As an improvement over the parabolic profile,  
we approximate the velocity profile 
by the cubic: 
\begin{equation} 
  u(r,z)/v(r) = a \eta + b \eta^2 + c \eta^3, 
\label{P3profile} 
\end{equation} 
where $a$, $b$, $c$ are now functions of $r$. 
Due to the boundary condition (\ref{rescbc}) and  
mass flux condition (\ref{avvelflux}), 
the coefficients $a$, $b$, and $c$ can be expressed  
in terms of one parameter $\lambda$ as, for example:    
\begin{equation} 
  a = \lambda + 3, \qquad 
  b = -(5 \lambda +3)/2, \qquad 
  c = 4 \lambda/3.  
\label{P3coefficients} 
\end{equation} 
The separation condition  
\begin{equation}  
  \left. u_z \right|_{z=0}=0  
\label{separcond}  
\end{equation}  
is now equivalent to $a=0$, or $\lambda=-3$. 
The $u$-profile is parabolic when $c=0$, or $\lambda=0$. 
 
Now that we have two unknowns $h(r)$ and $\lambda(r)$, 
two equations are necessary. 
We use the averaged momentum equation 
(\ref{averagedmomentum}) 
as the first equation. 
Note that $G$ is now not a constant, 
but depends on the shape parameter $\lambda$. 
From (\ref{squareaverage}), we obtain 
\begin{equation} 
  G(\lambda) = \frac{6}{5}- \frac{\lambda}{15}+ 
\frac{\lambda^2}{105} . 
\label{defF2} 
\end{equation} 
Following the Karman-Pohlhausen choice, 
we choose the second equation to be 
the momentum equation (\ref{rescBLrad}) 
evaluated at $z=0$: 
\begin{equation} 
  h' = \left. u_{zz} \right|_{z=0} . 
\label{momentumonbottom} 
\end{equation} 
This connects the pressure gradient on $z=0$ 
with $\lambda$. 
Using (\ref{P3coefficients}) and (\ref{defF2}), 
the two equations  
(\ref{averagedmomentum}) and (\ref{momentumonbottom}) can 
be  
written as 
\begin{equation} 
\begin{array}{l}  
 \displaystyle v \frac{d}{dr} 
 \left\{ G(\lambda) v \right\} = -h' - 
 \frac{v}{h^2}(\lambda+3) \\ 
 \displaystyle h'= -\frac{v}{h^2} (5 \lambda+ 3) 
\end{array} 
\label{choseneqns}  
\end{equation}  
which can be simplified to 
\begin{equation} 
  \begin{array}{l}  
    \displaystyle (G(\lambda) v)'= \frac{4\lambda}{h^2} \\ 
    \displaystyle h' = -v \frac{5\lambda +3}{h^2}.  
  \end{array}     
\label{shinya} 
\end{equation} 
Finally, eliminating $v$ using (\ref{avvelflux}), 
we obtain a nonautonomous system of two ordinary differential 
equations 
for $h(r)$ and $\lambda(r)$: 
\begin{equation} 
  \begin{array}{l}  
    \displaystyle h' = -\frac{5\lambda +3}{r h^3} \\ 
    \displaystyle \frac{d G}{d \lambda} \lambda' = 
      \frac{4 r \lambda}{h} + G(\lambda) 
      \frac{h^4 - (5 \lambda + 3)}{r h^4} . 
  \end{array}     
\label{shinya2} 
\end{equation} 
This is the model for the stationary circular hydraulic jump. 
It does become singular, 
but only on the lines $h=0$ and $\lambda=7/2$ 
which does not cause any trouble in describing a flow 
with a separated zone ($\lambda<-3$). 
We show in the next section that 
the highly simplified model indeed contains 
solutions which describe the observed circular hydraulic jumps. 
A similar approach using momentum and energy conservation 
was used in \cite{Arakeri}, but they did not succeed in 
finding continuous solutions through the jump. 
 
\subsection{Numerical solution of the integrated model} 
 
The model (\ref{shinya2}) can be solved as a boundary value 
problem  
by specifying two boundary conditions for different values of $r$.  
Thus we impose  
\begin{equation}  
  (r_1,h_1(r_1)) \quad \mbox{and} \quad (r_2,h_2(r_2)) , 
  \quad \mbox{$r_1<r_2$} 
\label{bvprob}  
\end{equation}  
where the values are taken from the measured surface height data. 
There is no fitting parameter once they are chosen, 
and the function $h(r)$ and the shape parameter $\lambda(r)$ 
are determined. 
In particular, we do {\em not} need to specify the shape parameter 
as a part of the boundary conditions. 
This is an advantage of the simplified model 
since one no longer needs to specify the velocity profile 
at the inlet and/or outlet boundaries, 
which is not easy to do. 
In fact, we see that specifying both $h$ and $\lambda$ at one $r$, 
either inside the jump or outside, and solving (\ref{shinya2}) 
as an initial value problem is unstable. 
The system is extremely sensitive to the initial condition 
if one integrates (\ref{shinya2}) 
in the direction of increasing $r$ from a small $r$ 
or in the direction of decreasing $r$ from a large $r$. 
Therefore, we choose $r_1$ and $r_2$ near 1, 
typically $r_1$ around 0.4-0.8 and 
$r_2$ around 1.2-1.6. 
Then, a straightforward shooting method from either boundary 
is sufficient to obtain a solution. 
After this is achieved, 
the solution is extended to $r<r_1$ and to $r>r_2$  
by integrating (\ref{shinya2}) backward from $r_1$ 
and forward from $r_2$, respectively. 
Integrations in these directions are stable. 
 
\begin{figure}[t] 
\centerline{\psfig{file=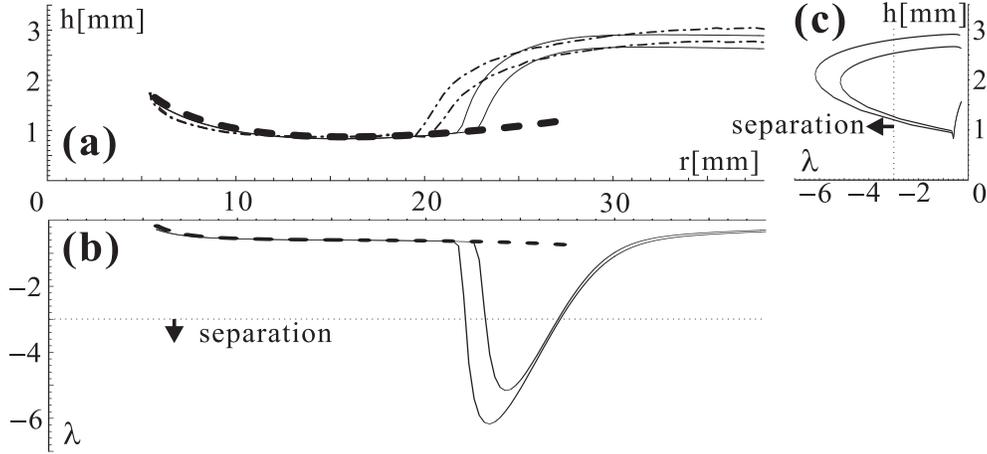,width=13cm}} 
\caption{ 
(a) Two surface height profiles of type I flow, 
taken from the experiment in (\protect{\ref{series4}}) 
are shown as the dot-dashed curves. 
Numerical solutions of the model (\protect{\ref{shinya}}) 
are shown as solid curves in both panels, 
and show reasonable agreement. 
To obtain each of the numerical solutions, 
$h$ values were read from the experimental data 
at $r=11.8$[mm] and $r=30.0$[mm], 
then a boundary value problem was solved by the shooting method. 
The thick dashed curve represents an analytical approximation 
of the solutions before the jump, 
described in Sec.\protect{\ref{sec:asymptotics}}.1. 
The formula (\protect{\ref{innerhas}}) 
and (\protect{\ref{innerphias}}) 
shows good agreement with one fitting parameter. 
(b) The computed shape parameters $\lambda(r)$, characterizing 
the velocity profiles, corresponding to the two numerical solutions 
in (a).
The flow is separated behind the jump where $\lambda<-3$, 
and approaches the parabolic profile $\lambda=0$ as $r$ increases. 
Again, the dashed curve is an analytical approximation. 
(c) Two trajectories of (\protect{\ref{shinya}}) are shown in the 
$(h,\lambda)$-plane. 
They correspond to solid curves in (a) and (b). 
} 
\label{compare1} 
\end{figure} 
 
Figure \ref{compare1}(a) shows two solutions 
of such a boundary value problem. 
They correspond to the two type I solutions 
in Fig.~\ref{series4}, 
reproduced here as dot-dashed curves. 
From each curve the boundary data are taken 
at $\tilde{r}_1 = 11.8$[mm] (corresponding to dimensionless value 
$r_1=0.42$) 
and $\tilde{r}_2 = 30.0$[mm] (to $r_2=1.07$). 
The computed solutions $h(r)$ corresponding to the data 
are shown in solid curves. 
Each curve shows a gradual decrease 
for small $\tilde{r}$ as $\tilde{r}$ increases, 
reaches a minimum at some $\tilde{r} \approx 15$[mm], 
and then undergoes a sharp jump at $\tilde{r} \approx 22$-
23[mm], 
and a slow decay after the jump. 
The location of the jump is about 10\% off in each case, 
and the slope behind the jump is noticeably different. 
However, the qualitative behavior is well captured 
by the simple model. 
Figure \ref{compare1}(b) shows the shape parameter $\lambda$. 
The velocity profile changes suddenly almost simultaneously with the 
rapid increase of the surface height, 
and a region where $\lambda<-3$, corresponding to separation, 
is observed in each case.\footnote{ 
If the downstream height is further reduced, however, 
the shape parameter $\lambda$ does not reach $\lambda = -3$, 
and there is no separated region. 
Thus, our model predicts that a (weaker) jump 
without an eddy is possible. 
The flow near the bottom still decelerates 
just after the jump.} 
The parameter $\lambda(r)$ recovers and appears to converge 
to $\lambda=0$ (the parabolic profile) as $r$ becomes large. 
 
\begin{figure}[t] 
\centerline{\psfig{file=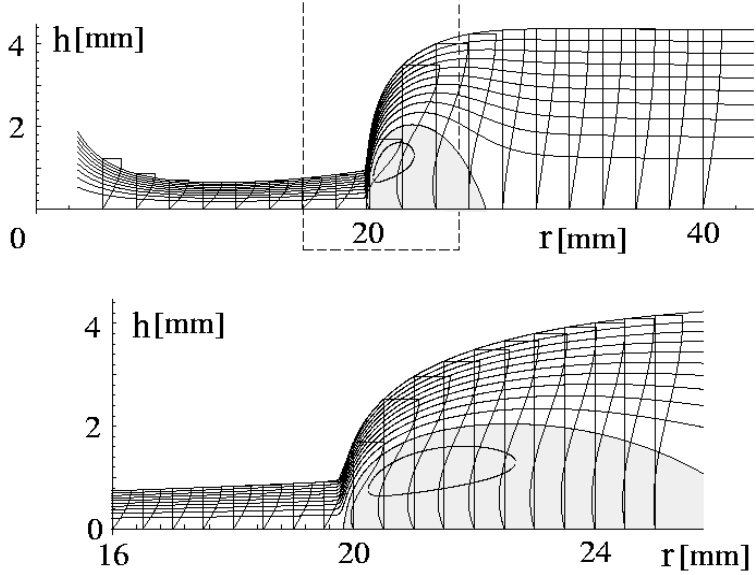,width=10cm}} 
\caption{ 
Visualization of the type I flow pattern 
based on the computed shape parameter $\lambda(r)$ from the 
model. 
The velocity profiles at equidistant locations in $r$ 
are the horizontal component $u$, 
thus they are not tangential to the streamlines. 
Since magnitudes of the velocity vary greatly 
between small and large $r$, 
the profiles of $u(r,z)/v(r)$ are shown. 
The streamlines separate zones which carry 10\% of the flow rate. 
A separation bubble is present in the range of $r$ where 
$\lambda<-3$. 
Note the difference in the scales for the axes. 
The parameters differ from those of 
Fig.~\protect{\ref{compare1}}. They are: 
$Q=33$[m$\ell$/s] and $\nu=1.4\times 10^{-5}$[m$^2$/s], 
corresponding to $r_* = 2.5$[cm], $z_* = 1.7$[mm], 
and $u_* = 16$[cm/s]. 
} 
\label{strfcn} 
\end{figure} 
 
The flow structure is more directly shown in Fig.~\ref{strfcn},
where the $u$-velocity profiles are computed from $\lambda$ 
at equidistant locations in $r$. 
Since magnitudes of the velocity vary a lot 
between small and large $r$, 
the profiles are scaled by the average velocity, 
so that the profiles of $u(r,z)/v(r)$ are shown. 
The stream function $\psi$ is computed from the definition 
\begin{equation} 
  u = \psi_z / r \quad , \quad 
  w = -\psi_r / r . 
\label{streamfunction} 
\end{equation} 
The dimensionless stream function varies from $\psi=0$ on $z=0$ 
to $\psi=1$ on $z=h$. 
Inside the separated region $\psi<0$.  
The contours at $\psi=-0.1,0,0.1\dots,1$ are shown in the figure. 
That is, a region between two neighboring contour curves 
carries 10\% of the mass flux. 
 
\begin{figure}[t] 
\centerline{\psfig{file=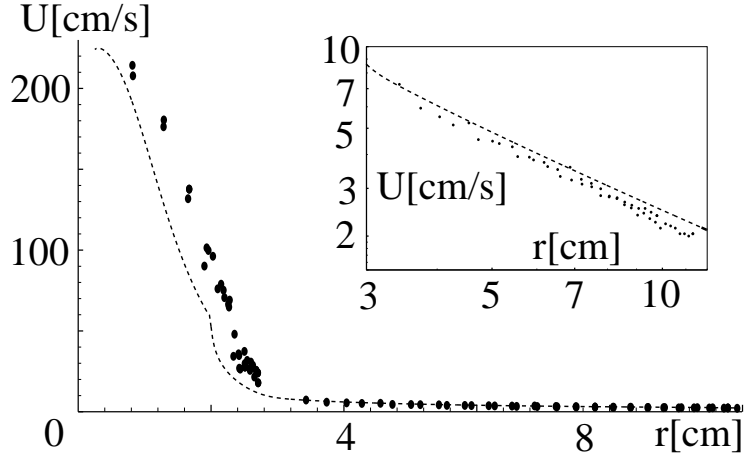,width=10cm}} 
\caption{ 
Comparison of the prediction from the model with 
a surface velocity measurement 
by C.\ Ellegaard, A.E.\ Hansen,  
and A.\ Haaning \protect{\cite{behh}}. 
The parameters are the same as in Fig.~\protect{\ref{strfcn}}. 
Marker particles and a high-speed camera were used 
in order to obtain the surface velocity $U$ shown as dots. 
The theoretical dotted curve was computed by finding a stationary 
solution $h(r)$ and $\lambda(r)$ of a boundary value problem 
using two data points taken from the measured surface profile (not 
shown). 
Although the location of the jump is about 20\% off, 
the model reproduces qualitative feature of the measurement very 
well. 
At small $r$, the velocity drops rapidly and almost linearly. 
It then shows a cusp-like drop at the jump, 
and decays gradually for large $r$. 
The final decay is proportional to $1/r$ as can be seen 
from the slope of about $-1$  
in the log-log plot of the exterior region (inset). 
} 
\label{compare2} 
\end{figure} 
 
The surface velocity $U$ predicted from the model 
is shown in Fig.~\ref{compare2}. 
The parameters are the ones used in Fig.~\ref{strfcn}. 
The model again misses the location of the jump by about 20\%, 
so measurements and the curve from the model are offset, 
but qualitative features are well reproduced. 
The velocity outside the jump is small and 
decays like $U \propto 1/r$, 
as can be seen from the log-log plot in the inset. 
This is consistent with an almost constant $h$ 
and a nearly parabolic velocity profile, 
which we analytically demonstrate in the next section. 
On the other hand, the surface velocity decreases 
almost linearly before the jump. 
This region is harder to explain intuitively, 
but an analytical approximation is also obtained in the next section. 
At the jump a rapid, cusp-like drop in the velocity is noticed. 
 
Finally, we discuss the dependence of the solutions 
on the external height $h_{ext}$. 
Both in experiments and in the model 
the height inside the jump is little affected 
by the change in the external boundary condition $h_2(r_2)$. 
The numerical solutions as well as the measured 
surface profiles in Fig.~\ref{compare1}(a,b) 
apparently overlap in the interior to the jump.    
Of course, the two solutions must 
correspond to different trajectories 
of the model (\ref{shinya}) and 
cannot collapse exactly onto a single curve. 
However, the closeness of the solution curves 
in the interior to the jump  
is the cause of the difficulty of solving the initial value 
problem starting from a small $r$. 
 
If the external height is further increased, 
a transition from type I to II is observed in the experiment, 
as illustrated  in Fig.~\ref{series4} and Fig.~\ref{type1-2}. 
Unfortunately, no such transition is reproduced 
in the model when $h_2$ is increased.  
Instead, one finds a computed solution of the model 
similar to the ones in Fig.~\ref{compare1} 
even for a much larger $h_2$. 
A physical mechanism to ``break'' the wave into a type II flow 
appears to be missing. 
In fact, a solution with a roller is prohibited 
by the model (\ref{shinya}).  
The surface velocity on a roller is negative (inward). 
According to (\ref{P3coefficients}), 
the velocity at the surface is  
\begin{equation}  
  U=v (a + b +c) =v \frac{9- \lambda}{3},    
\label{topvel}  
\end{equation}  
where $v>0$  is the average velocity.  
Thus, $U<0$ iff $\lambda>9$. 
However, since we start with $\lambda \simeq 0$ and   
the line $\lambda=7/2$ makes (\ref{shinya}) singular, 
a solution with a roller is not possible. 
It seems likely that this behavior can be traced back to 
the assumed pressure distribution (\ref{hydrostaticpressure}) 
which does not provide any pressure gradient 
along the surface $z=h$. 
In a recent simulation of the circular hydraulic jump 
by Yokoi {\em et al.}\cite{Yokoi} 
pressure buildup just behind the jump is observed 
and claimed to be crucial in breaking the jump. 
The non-hydrostatic pressure arises partly due to 
the surface tension in (\ref{dynamicbc}.1), 
but also due to the truncated viscous terms in 
(\ref{nondimBLeeq0}) 
and (\ref{dynamicbc}). 
We do not know at present how best to extend our model 
to include the type II flows. 
 
\subsection{Asymptotic analysis of the averaged system} 
\label{sec:asymptotics} 
 
In this section we approximate the solutions 
of (\ref{shinya}) analytically 
using formal perturbation expansions. 
We obtain explicit expressions for two ``outer'' regions: 
the region before the jump and the one after the jump. 
Moreover, we derive a single ordinary differential equation 
for the ``inner'' region near the jump. 
Analysis in the inner region 
connects a previous model using a Rayleigh shock 
with our model. 
 
\subsubsection{Outer solution 1 (before the jump)} 
 
First, we analyse the region before the jump 
where thickness of the fluid as well as the radius 
are small, compared to the exterior region. 
We denote the typical thickness, in the dimensionless coordinates, 
as $\theta$, and treat it as a formal small parameter. 
We rescale the variables into $H$, $R$, and $V$ as 
\begin{equation}  
  \begin{array}{l}  
    h= \theta H , \\  
    r= \theta^ \alpha R , \\  
    v= \theta^{-1-\alpha} V , 
  \end{array}  
\label{asymp_resc_inner}  
\end{equation}  
and require consistent balance of the terms in (\ref{shinya2}) 
or, equivalently, (\ref{shinya}). 
The rescaling for $v$ in the third equation of 
(\ref{asymp_resc_inner}), 
is chosen to ensure  mass conservation (\ref{avvelflux}) 
for all  $\theta$. 
In terms of the new variables, (\ref{shinya}) can be written as 
\begin{equation} 
  \begin{array}{l}  
    \theta^{-2 \alpha -1} \displaystyle  
      \frac{d}{dR} \left( G(\lambda) V \right) =  
    \theta^{-2}\frac{4\lambda}{H^2} , \\ 
    \theta^{1- \alpha }\displaystyle H' = - \theta^{- \alpha-3 } 
    V \frac{5\lambda +3}{H^2}.  
  \end{array}     
\label{shinya_resc_inner} 
\end{equation} 
From the first equation the only consistent choice 
is to take $\alpha=1/2$. 
Then, in order to balance the power of $\theta$ 
on both sides of the second equation, 
we need 
\begin{equation}  
\lambda = -3/5 + \theta^4 \lambda_1 + \ldots . 
\label{lambda_asympt_inner}  
\end{equation} 
The form is also motivated by Fig.~\ref{compare1} 
in which $\lambda$ stays close to the value $-0.6$ 
before the jump. 
 
To find $H(R)$ and the correction $\lambda_1$, 
substitute (\ref{lambda_asympt_inner}) into 
the first equation of (\ref{shinya_resc_inner}). 
To the lowest order in $\theta$ we obtain 
\begin{equation}   
  G(-0.6) \left( \frac{1}{H^2 R} \frac{dH}{dR} + \frac{1}{H 
R^2} \right)= 
  \frac{12}{5 H^2},  
\label{momeqint} 
\end{equation} 
where $G(-0.6) = 1088/875 \simeq 1.243$. 
Solving this equation yields 
\begin{equation}  
  H= \frac{C_1}{R} + \frac{4}{5 G(-0.6)} R^2,  
\label{innerhas}  
\end{equation}  
where $C_1$ is an arbitrary integration constant. 
The functional form agrees  
with Watson's self-similar solutions \cite{Watson}. 
We also compare the lowest order term of $\theta$ 
in the second equation of (\ref{shinya_resc_inner}), 
and find that  
\[ 
  \lambda_1= \frac{R H^3}{5} \frac{dH}{dR} . 
\] 
By substituting $H$ in (\ref{innerhas}) we obtain 
an approximate expression for $\lambda$:   
\begin{equation}  
  \lambda=-\frac{3}{5}+\theta^4 \left[ 
    \frac{H^4}{5} - \frac{12}{25 G(-0.6)} R^2 H^3 \right] .  
\label{innerphias} 
\end{equation}  
We test the approximations (\ref{innerhas}) and (\ref{innerphias}) 
in Fig.~(\ref{compare1}). 
The dashed curves are the theoretical curves of $H(R)$ and 
$\lambda(R)$, 
shown in the dimensional coordinates. 
They match the numerical solutions and the measurements 
well before the jump. 
Here, the formal parameter $\theta$ is taken as unity, 
and the one free parameter $C_1$ was fitted to be $0.25$. 
 
\subsubsection{Outer solution 2 (after the jump)} 
 
Let us now consider the behavior of (\ref{shinya}) for large $r$. 
We again introduce a formal small parameter $\theta$, 
but we now rescale $r= \theta^{-1} R$.  
If we moreover assume that the height is of order $1$, i.e., $h=H$, 
then the rescaling of the velocity is necessarily $v=\theta V$ 
due to (\ref{dimflux}).  
Using these new variables, Eqs.~(\ref{shinya}) become:   
\begin{equation} 
\begin{array}{l}  
  \theta^{2} \displaystyle \frac{d}{dR} \left( G(\lambda) V \right) 
= 
     \frac{4\lambda}{H^2} \\ 
  \displaystyle \frac{dH}{dR} = - V\frac{ 5 \lambda + 3}{H^2}.  
\end{array}     
\label{shinya_resc_outer} 
\end{equation} 
In order to balance the terms in the first equation we choose 
\begin{equation}  
  \lambda= \theta^2 \lambda_1 + \ldots . 
\label{lambdaresc} 
\end{equation}  
This is again consistent with the bottom panel  
of Fig.~\ref{compare1} where  
$\lambda$ apparently tends to $0$, 
corresponding to the parabolic profile. 
Then, the terms of order unity in the second equation are 
\begin{equation}  
  \frac{dH}{dR}=-\frac{3}{R H^3} 
\label{outerhas} 
\end{equation}  
whose solution is 
\begin{equation}  
  H = \left( 12 \log \frac{R_{\mbox{\scriptsize end}}}{R} 
\right)^{1/4} 
\label{singheight}  
\end{equation}  
where $R_{\mbox{\scriptsize end}}$ is an integration constant 
representing the radius where the height goes to $0$. 
Thus, (\ref{shinya}), as well as the simpler Kurihara-Tani  
model (\ref{Tanieqn}), becomes singular when $r \rightarrow 
\infty$.  
This seems to be a general property of  models based on 
the boundary layer equations \cite{BDP}. 
The absence of regular solutions for the system  
(\ref{rescBLrad})-(\ref{rescbc}) when $r \rightarrow \infty$ was 
proved  
 in  \cite{HHRputk}. We have attributed   
this lack of asymptotic solutions  
to the influence of the finite size of the plate.  
Indeed, a solution with vanishing height such as  
(\ref{singheight}) reminds one very  
much of a flow running off the edge of a circular plate.    
 
The height $H(R)$, given by equation (\ref{singheight}),  is a very 
slowly  
varying function of $R$. There is a long regime 
$1 \ll R \ll R_{\mbox{\scriptsize end}}$ 
where the height appears almost constant. 
In this intermediate regime 
the leading order of (\ref{shinya_resc_outer}.1) becomes 
\begin{equation}  
  G(0) \frac{d}{dR} \left( \frac{1}{R H} \right) =  
  \frac{4 \lambda_1}{H^2} 
\label{outerphias}  
\end{equation}  
where $G(0)=6/5$. 
Therefore, 
\begin{equation} 
\lambda=\theta^2  \lambda_1 = -\theta^2 \frac{G(0) H^2}{4} \left(  
    \frac{1}{RH^2} \frac{dH}{dR} + \frac{1}{R^2 H} \right) 
  \approx \frac{3}{10 r^2} \left( \frac{3}{H^3} - H \right) . 
\label{daarligt} 
\end{equation} 
We conclude that $\lambda(R) \propto 1/R^2 \rightarrow 0$ 
which explains the observed approach 
to the parabolic velocity profile for large $r$. 
 
\subsubsection{Inner solution near the jump: conservation of 
momentum}  
 
Finally, we analyze the region around the hydraulic jump. 
Recall that in the Kurihara-Tani theory (\ref{Tanieqn})  
the jump was obtained by fitting a Rayleigh shock. 
In this section, we show that our model (\ref{shinya2})  
is a natural generalization of the equation. 
 
To do this we return to (\ref{choseneqns}), 
and introduce a formal parameter $\mu$ 
in the left-hand side of the second equation. 
\begin{equation} 
\begin{array}{l}  
 \displaystyle v \frac{d}{dr} 
 \left\{ G(\lambda) v \right\} = -h' - 
 \frac{v}{h^2}(\lambda+3) \\ 
 \displaystyle \mu h'= -\frac{1}{r h^3} (5 \lambda+ 3). 
\end{array} 
\label{mueq}  
\end{equation}  
where $v=1/(rh)$. 
The first equation describes the balance of inertia, 
hydrostatic pressure, and viscous forces. 
The value $\mu=1$ corresponds to (\ref{choseneqns}). 
 
Setting $\mu=0$ gives $\lambda=-0.6$. 
Then the first equation becomes the Kurihara-Tani equation 
(\ref{Tanieqn}), 
except that the coefficient $6/5=1.2$ is changed to 
$G(-0.6) \approx 1.243$ here, 
since the profile is not parabolic. 
(As discussed before, the velocity profile 
is not so important in their model as long as it is self-similar.) 
Since our model corresponds to $\mu =1$, 
the parameter $\mu$ interpolates between the two models, 
but the correspondence of the two is not obvious 
because the limit $\mu=0$ is a singular limit. 
We treat $\mu$ as a formal small parameter, 
and carry out a singular perturbation analysis 
to investigate the connection 
as well as to obtain an approximation in the jump region. 
 
In Kurihara-Tani model a shock is needed  
to extend the solution from small to large values of $r$. 
Suppose the shock is situated at $r=r_0$. 
Consider a small region of size $\mu$ around $r=r_0$, 
and rescale the coordinate as $r=r_0 + \mu X$. 
Then, in the inner coordinate $X$, Eq.~(\ref{mueq}) becomes   
\begin{equation} 
 \begin{array}{l}  
  \displaystyle \frac{1}{r_0 h} \frac{d}{dX} \left\{ 
    \frac{G(\lambda)}{r_0 h} \right\} = 
     -\frac{dh}{dX} +O( \mu ) \\ 
  \displaystyle \frac{dh}{dX} = - \frac{5 \lambda + 3}{r_0 h^3} + 
O(\mu) .  
\end{array} 
\label{asmueq}  
\end{equation} 
We see that $\lambda=-0.6$ with $h$ an arbitrary constant
are the only possible fixed points of (\ref{asmueq}).
Thus the solutions must satisfy $\lambda \rightarrow -0.6$
for $X \rightarrow \pm
\infty$. This correctly matches the external solution before
the jump, but not after the jump, where $\lambda \rightarrow 0$.\footnote{ 
Note that the singularity of the outer solution after the jump 
(\ref{singheight})-(\ref{daarligt}) for $r \rightarrow 0$ does not 
allow correct matching for $X \rightarrow + \infty$ when $\mu \rightarrow 0$. 
Nevertheless, our method reproduces the structure of the separation 
zone quite well. }
The first equation can be integrated once, 
giving the momentum conservation. 
\begin{equation}  
  \frac{G (\lambda)}{r_0^2 h} + \frac{h^2}{2} = C_3 
\label{momcons} 
\end{equation}   
with an integration constant $C_3$. 
Now we solve the second equation of (\ref{asmueq}) for 
$\lambda$, 
and substitute it into this equation. 
Using (\ref{defF2}) in the form 
$G (\lambda) = \frac{1}{105} \left( \lambda - \frac{7}{2} 
\right)^2 
+ \frac{13}{12}$, 
we obtain an ordinary differential equation for $h$ only: 
\begin{equation} 
  \frac{1}{105} \left( \frac{r_0 h^3}{5} \frac{dh}{dX}  
    + \frac{41}{10} \right)^2 + \frac{13}{12} + \frac{r_0^2 
h^3}{2} 
  = C_3 r_0^2 h . 
\label{middleheq0}   
\end{equation} 
We look for a solution $h(X)$ with 
$h \rightarrow h_1$ as $X \rightarrow -\infty$ 
and $h \rightarrow h_2$ as $X \rightarrow +\infty$ 
where $h_1$ and $h_2$ are constants. 
Then, Eq.~(\ref{middleheq0}) with the first boundary condition 
determines 
the constant $C_3$ in terms of $r_0$ and $h_1$. 
Eliminating $C_3$ we obtain 
\begin{equation} 
 \frac{1}{105} \left[ \left( \frac{r_0 h^3}{5} \frac{dh}{dX} 
    + \frac{41}{10} \right)^2 h_1 - \left( \frac{41}{10} \right)^2 h 
 \right] + \frac{13}{12} (h_1-h) - \frac{r_0^2}{2} h_1 h ( h_1^2-
h^2 ) = 0 . 
\label{middleheq}   
\end{equation}  
Plugging the second boundary condition into this equation 
yields a relation between $h_1$ and $h_2$, given $r_0$. 
\begin{equation}  
  h_1 h_2^2 + h_1^2 h_2 - 2 h_c^3 = 0 . 
\label{analogousRayleigh} 
\end{equation} 
where 
\begin{equation} 
  h_c = ( G(-0.6)/r_0^2 )^{1/3} 
\label{hcr}  
\end{equation} 
is the critical height for the 
circular hydraulic jump.\footnote{ 
In dimensional variables,
the critical height is
$\tilde{h}_c = ( G(-0.6) q^2 / g \tilde{r}_0^2 )^{1/3}$. 
This is identical to the critical height  
(\protect{\ref{criticalheight}}) that 
appeared in the Rayleigh shock, apart from the numerical factor 
and the influence of $\tilde{r}_0$ reflecting the radial geometry. 
The viscosity $\nu$ only enters in the coefficient 
of $dh/dX$ in the dimensional version of (\ref{middleheq}), 
thus does not affect $\tilde{h}_c$.} 
Solving this equation, we obtain 
an equation analogous to the shock condition (\ref{Froude}): 
\begin{equation}  
  \frac{h_2}{h_1} = \frac{1}{2} \left( 
     -1 + \sqrt{1+8 (h_c/h_1)^3} \right)  
    = \frac{2}{-1 + \sqrt{1+8 (h_c/h_2)^3}} . 
\label{h2sol}  
\end{equation} 
It is easy to see that $h_c$ is always between $h_1$ and $h_2$, 
i.e., $h_1 < h_c < h_2$ or $h_2 < h_c < h_1$. 
The Froude number in this case could naturally be defined 
as $F(X)^2 = ( h_c/h(X) )^3$.\footnote{ 
However, it is not clear whether $F$ defined in this way 
can be a measure of super- and subcriticality since 
the governing equations are not the shallow water equations 
and therefore propagation of disturbances do not obey 
the well-known velocity $\sqrt{gh}$.} 
 
When $h_1$ is close to $h_c$, 
the final height $h_2$ is close to $h_c$ as well. 
Then, the Froude number is close to unity for all $X$, 
and the jump is weak, i.e.\ $h_c - h_1 = \delta \ll 1$. 
Then, we see from the balance of the terms in (\ref{middleheq}) 
that 
$h = h_c + \delta Y (\delta x)$. 
The leading balance reduces to 
\[ 
  Y' = \gamma ( 1 - Y^2 ) 
\] 
with 
\begin{equation}  
  \gamma=\frac{196875}{1312} \left( \frac{7}{17} \right)^{2/3} 
  r_0^{5/3} \approx 83.1 r_0^{5/3} . 
\label{expgr}  
\end{equation} 
Thus, in the weak jump limit, the height is given by 
\begin{equation}  
  h(x) = h_c + \delta \tanh ( \delta \gamma x ) . 
\label{smalljumpsol} 
\end{equation}  
 
\begin{figure} 
\centerline{\psfig{file=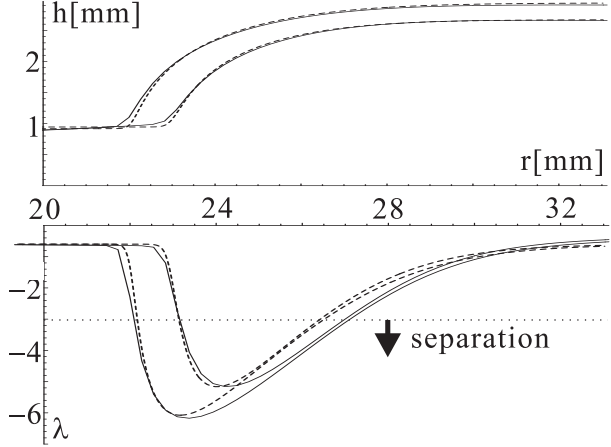,width=8cm}} 
\caption{  
Comparison between the full numerical solution of 
(\protect{\ref{shinya}}), 
the same two solutions as in Fig.~\protect{\ref{compare1}} 
shown as solid curves, 
and solutions of the asymptotic equation 
(\protect{\ref{middleheq}}), 
shown as dashed curves. 
Even though the asymptotic analysis assumes 
$\mu \rightarrow 0$, 
the solutions compare fairly well with the full numerics 
corresponding to $\mu=1$. 
The asymptotic analysis connects the model 
(\protect{\ref{shinya}}) 
with the Rayleigh shock condition.  See text. 
} 
\label{asympheight} 
\end{figure} 
 
It is interesting to note that  
we can connect from $h_1$ at $X=-\infty$ to $h_2$ at $X=+\infty$ 
if $h_1 < h_2$, 
but not if $h_1 > h_2$, 
just like in the Rayleigh shock. 
This requirement comes from  
the equation (\ref{middleheq}) self-consistently 
rather than making a hypothesis 
on the energy loss like we did in (\ref{dissenr}). 
To see this, consider the stability 
of the fixed points $h_1$ and $h_2$ 
with respect to the governing equation (\ref{middleheq}) 
for $h$.\footnote{Of 
course, this stability analysis is to study 
existence of stationary solutions, 
and not to study the stability of such solutions 
in the time-dependent theory.} 
Linearizing (\ref{middleheq}) around the uniform solutions 
$h_i$ (where $i=1,2$), 
we obtain an equation for the perturbation $\delta h_i$ in the 
height: 
\[ 
  \frac{d}{dX} (\delta h_i) = K_i \delta h_i 
\] 
where  
\begin{equation}  
  K_i = \frac{2625}{41} 
  \frac{r_0 \{ 2 h_c^3 + h_1 (h_1^2-3 h_i^2)\} }{2 h_i^3 h_1} . 
\label{linstab}  
\end{equation} 
If $h_1 < h_c < h_2$, then $K_1 > 0 > K_2$, 
showing that the fixed point $h=h_1$ is unstable and $h=h_2$ 
stable. 
A trajectory departing from $h_1$ 
at $X=-\infty$ and arriving at $h_2$ at $X=+\infty$ 
is not prohibited, and we can indeed find such a trajectory 
shown in Fig.~\ref{asympheight}. 
In contrast, if $h_1 > h_c > h_2$, then 
the stability of the fixed points is reversed, 
and there is no trajectory going from $h_1$ to $h_2$. 
 
When $h_1 < h_c < h_2$ so that such a trajectory exists, 
the departure from $h_1$ is generally rapid, 
giving an impression of a ``sharp corner'' at the beginning of the 
jump, 
and the arrival at $h_2$ is much smoother 
just as shown in Fig.~\ref{asympheight}. 
This is because the magnitude of the stability coefficient $K_1$ 
is large compared to that of $K_2$. 
The feature is most pronounced when $h_1$ is small (so, $h_2$ is 
large). 
It vanishes as $(h_2 - h_1) \rightarrow 0$ 
when $K_1$ and $K_2$ both tend to zero. 
 
In Fig.~\ref{asympheight} 
we compare solutions of (\ref{middleheq}) 
with the two solutions of the full numerical solution 
of (\ref{shinya}) shown in Fig.~\ref{compare1}. 
The jump region is enlarged. 
Solutions of (\ref{middleheq}), 
shown as solid curves, 
are computed by fitting the values 
for $h_1$ and $h_2$, and solving the equation using $r_0$ 
obtained from (\ref{analogousRayleigh}) and (\ref{hcr}). 
We chose an initial condition to be somewhere inside the jump, 
and integrated (\ref{middleheq}) forward and backward from it. 
Since (\ref{middleheq}) has a translational invariance 
with respect to $X$, 
the initial condition fixes the location of the jump 
without affecting the shapes of $h$ or $\lambda$. 
The analysis assuming $\mu \rightarrow 0$ performs surprisingly 
well 
against the numerical solution for $\mu=1$. 
The size of the jump region is now of order $\mu$, i.e., unity, 
and the internal structure is non-trivial. 
The single ordinary equation (\ref{middleheq}) is capable 
of describing the eddy formation in this region. 
   
%%%%%%%%%%%%%%%%%%%%%%%%%%%%%%%%%%%%%%%%%%%%%%%%%%%%%%%%%%%%%%%%%%%%%%%%% 
\clearpage 
\section{Flow down an inclined plane} 
 
\subsection{Introduction to the problem} 
 
The properties of waves running down an inclined 
plane is a subject of great theoretical and practical importance, 
and has attracted the attention of many researchers.  
Starting with the pioneering work of  
Kapitsa \& Kapitsa \cite{Kapitsa}, some of the 
major contributions  
to this field  are found 
in 
\cite{Benjamin,Benney,Nakaya,Pumir,Chang1,Chang2,Gollub,Lee}
.  
The physical picture is the following. 
A fixed flux of fluid is constantly poured 
onto the inclined plane from above. 
The fluid forms a stream moving  
downwards under the action of gravity -- an idealized model of a 
river. 
If the influx of fluid upstream is suddenly increased,
it  causes the height upstream to increase, 
and the extra mass of fluid to propagate downstream. In a river, 
this may be caused by the melting of snow  at regions neighbouring 
the river's source, or by sudden rain. 
A river bore, on the other hand, is introduced at the mouth of  
the river by a tidal wave,  
for instance, and moves upstream.  
In both cases, a solitary wave can be formed, 
moving at a constant velocity $c$ without changing its shape. 
 
We are particularly interested in kink-like solitary wave solutions 
going from one constant height $h_1$ to another $h_2$. 
One can identify such a solution with a heteroclinic orbit, 
connecting two stationary states \cite{Pumir}. 
The speed $c$ depends on how much the fluid level is increased, 
i.e., the heights $h_1$ and $h_2$. 
Alternatively, we can consider $c$ as a parameter, and study  
the existence of the stationary solution 
$h \equiv \mbox{const.}$ depending on $c$.  
It is rather straightforward to see that two solutions 
with $h \equiv h_1$ and $h \equiv h_2$ exist 
if $c$ is sufficiently large. 
However, even if $c$ is in that regime, 
it is hard to judge whether there exists a smooth  
solution connecting the two states.   
 Based on the method of averaging in Sec.~2, 
we develop a simple model 
which helps us to derive criteria for their existence 
and to compute the wave form. 
The model also enables us to ask 
whether they appear as ``Rayleigh shocks'' 
in the sense that the flow is supercritical  
in front of the kink structure and subcritical behind it. 
As we shall elaborate, 
the distinction between super- and subcritical flows is 
a concept inherent in inviscid shallow water theory, 
and is not at all obvious for a viscous flow 
since now the waves will show dispersion as well as damping. 
Indeed, we find that  
the wave velocities corresponding to the largest wave lengths  
will always propagate both forward and backwards, 
as in a subcritical flow. 
Nevertheless, 
if we focus on wavelengths of the order of the depth of the fluid 
layer,  
a clear distinction can be made. 
 
There is another kind of flow in the linear geometry 
in which a sudden thickening of height is observed. 
This solution is not only relevant for, e.g., 
the flow of water exiting from a sluice 
but is also a direct analog of the circular hydraulic jump. 
The flow streams rapidly in a region immediately after the sluice, 
and then abruptly slows down at a certain downstream position. 
It is stationary (i.e.\ $c=0$) with a constant discharge, 
and is {\em not} obtained as a state connecting 
two ``equilibrium'' heights. 
In fact, the rapid flow before the jump cannot be extended 
arbitrarily far upstream.
We shall show that our models provide  
physically reasonable solutions in this case, too. 
 
In Secs.~3.2 and 3.3 we write down the complete system 
for the inclined plane problem, non-dimensionalize it, 
simplify it using the boundary layer approximation, 
and average over the thickness in two ways. 
These steps are in parallel with those in Sec.~2, 
but we go through them briefly not only for completeness 
but also since the geometry and the characteristic scales 
are different. 
To seek stationary and traveling wave solutions, 
we write the equations in a coordinate frame  
moving at a constant speed in Sec.~3.4. 
Traveling waves are studied in detail in Sec.~3.5, 
and the stationary jumps in Sec.~3.6. 
 
\subsection{The governing equations} 
 
We consider a viscous, incompressible, two-dimensional flow. 
The coordinate system is  
$\tilde{x}$ in the downstream direction 
parallel to the inclined plane, 
and $\tilde{y}$ in the perpendicular direction above the plate. 
Denote the velocities in these directions by  
$\tilde{u}(\tilde{x},\tilde{y},\tilde{t})$ and  
$\tilde{w}(\tilde{x},\tilde{y},\tilde{t})$, respectively, 
the pressure by $\tilde{p}(\tilde{x},\tilde{y},\tilde{t})$, 
and the height by $\tilde{h}(\tilde{x},\tilde{t})$. 
The governing equations for this problem are 
the continuity equation 
\begin{equation} 
  \tilde{u}_{\tilde{x}} + \tilde{w}_{\tilde{y}} = 0  
\label{xycontinuity} 
\end{equation} 
and the Navier-Stokes equations 
\begin{equation} 
  \begin{array}{l}  
    \displaystyle 
    \tilde{u}_{\tilde{t}} +  
    \tilde{u} \tilde{u}_{\tilde{x}} +  
      \tilde{w} \tilde{u}_{\tilde{y}} =  
     -\frac{1}{\rho} \tilde{p}_{\tilde{x}} +g \sin \alpha 
     +\nu \left( \tilde{u}_{\tilde{x} \tilde{x}} 
       +\tilde{u}_{\tilde{y} \tilde{y}} \right) \\  
    \displaystyle 
    \tilde{w}_{\tilde{t}} +  
    \tilde{u} \tilde{w}_{\tilde{x}} 
     +\tilde{w} \tilde{w}_{\tilde{y}} =  
     -\frac{1}{\rho} \tilde{p}_{\tilde{y}} -g \cos \alpha 
     +\nu \left( \tilde{w}_{\tilde{x} \tilde{x}} 
       +\tilde{w}_{\tilde{y} \tilde{y}} \right) 
  \end{array}  
\label{fullns} 
\end{equation} 
Here, $\alpha$ is the angle of the inclined plane 
(between 0 and $\pi/2$) 
measured downward from the horizontal line, 
and the subscripts denote the partial derivatives as before. 
The boundary conditions are identical to those of the radial 
geometry, 
i.e., (\ref{no-slipdim})--(\ref{kinematicbcdim}), 
by reading $\tilde{r}$ as $\tilde{x}$ and $\tilde{z}$ as 
$\tilde{y}$. 
The  local mass flux is:
\[ 
  \tilde{q}(\tilde{x},\tilde{t}) =  
  \int_0^{\tilde{h}(\tilde{x},\tilde{t})} 
  \tilde{u} d\tilde{y} . 
\] 
Integrating the continuity equation (\ref{xycontinuity}) in 
$\tilde{y}$ 
over the thickness and using the boundary conditions, 
we obtain the flux conservation equation: 
\begin{equation} 
  \tilde{h}_{\tilde{t}} + \tilde{q}_{\tilde{x}} = 0 . 
\label{fluxconservation} 
\end{equation} 
The equations above form a complete system 
apart from the inlet and outlet conditions. 
They possess a trivial stationary solution 
(Nusselt solution) 
with a constant $\tilde{h}$ and the parabolic velocity profile: 
\begin{equation}  
  \tilde{u}(\tilde{x},\tilde{y},\tilde{t}) \equiv  
  \frac{g \sin \alpha}{\nu} \left(\eta - \frac{\eta^2}{2}\right),  
\label{parab} 
\end{equation} 
where $\eta = \tilde{y}/\tilde{h}$. 
Given this equilibrium flow, 
the local flux $\tilde{q}$ is also uniform and steady, 
and is a function of $\tilde{h}$: 
\begin{equation}  
  \tilde{q} = \int_0^{\tilde{h}} 
    \tilde{u} d\tilde{y} = \frac{g \tilde{h}^3 \sin \alpha}{3 \nu} . 
\label{fluxpar} 
\end{equation} 
 
In a non-equilibrium flow we assume that the inclined plane 
is infinitely long, 
and the flow sufficiently far  downstream approaches this 
equilibrium flow. 
We then treat the flow rate $\tilde{q}$  
for $\tilde{x} \rightarrow \infty$ as the characteristic 
mass flux $q_*$. 
The corresponding height $\tilde{h}$ using (\ref{fluxpar}) 
is used as the length scale $h_*$, 
and $v_* = q_*/h_*$ becomes the characteristic velocity. 
We non-dimensionalize the governing equations by these scales. 
The continuity equation is unchanged in form: 
\begin{equation}  
  u_x + w_y = 0 , 
\label{resccontinuity} 
\end{equation} 
and the Navier-Stokes equations become 
\begin{equation}  
 \begin{array}{l}  
  \displaystyle 
  u_t + u u_x + w u_y = - p_x + \frac{3}{R} 
    + \frac{1}{R} ( u_{xx} + u_{yy} ) \\  
  \displaystyle 
  w_t + u w_x + w w_y = - p_y - \frac{3}{R \tan \alpha} 
    + \frac{1}{R} ( w_{xx} + w_{yy} ) 
 \end{array}  
\label{rescns}  
\end{equation} 
where the pressure is normalized to $\rho u_*^2$, 
and the Reynolds number is  
\begin{equation}  
  R = \frac{v_* h_*}{\nu} = \frac{q_*}{\nu} = 
     \frac{g {h_*}^3 \sin \alpha}{3 \nu^2} . 
\label{rnumincpl}  
\end{equation}  
The dimensionless mass flux is $q(x,t) = h v$ 
in terms of the average velocity 
\begin{equation}  
  v(x,t) = \frac{1}{h} \int^h_0 u dy  
\label{avvelip}  
\end{equation}  
whereby (\ref{fluxpar}) becomes 
\begin{equation} 
  q = hv = h^3 
\label{ndfluxpar} 
\end{equation} 
in an equilibrium flow of height $h$. 
 
\subsection{Boundary layer equations and averaged models} 
 
Since the flow on the inclined plane is expected to be 
predominantly 
in the $x$-direction, 
the boundary layer approximation should be applicable 
\cite{Chang1,Chang2} 
as long as separation does not occur.
In a similar manner as the radial case, 
the dominant terms of (\ref{rescns}) are: 
\begin{equation}  
 \begin{array}{l}  
  \displaystyle 
  u_t + u u_x + w u_y= - p_x + \frac{3}{R} + \frac{1}{R} u_{yy} 
\\ 
  \displaystyle 
  0 = -p_y - \frac{3}{R \tan \alpha} . 
\end{array}  
\label{BLeqinpl}  
\end{equation} 
The dynamic boundary conditions on $z=h$ reduce, as before, to: 
\begin{equation} 
 \begin{array}{l} 
  \left. p \right|_{y=h} = W h_{xx} \\ 
  \left. u_y \right|_{y=h} = 0 
 \end{array} 
\label{xyblbc} 
\end{equation} 
with the Weber number in this case being 
\begin{equation} 
  W = \frac{\sigma}{\rho h_* {v_*}^2} = 
  \frac{9 \sigma}{\rho g {h_*}^2 \sin^2 \alpha} . 
\label{webernum} 
\end{equation}  
From (\ref{BLeqinpl}.2) and (\ref{xyblbc}), the pressure is 
hydrostatic with contribution from the surface tension: 
\begin{equation} 
  p(x,y,t) = \frac{3}{R \tan \alpha} \left( h(x,t)-y \right) + W 
h_{xx} 
\label{xyhydrostaticpressure} 
\end{equation} 
so, (\ref{BLeqinpl}.1) becomes 
\begin{equation}  
  u_t + u u_x + w u_y = \frac{3}{R} - \frac{3}{R \tan \alpha} h_x 
    + \frac{1}{R} u_{yy} + W h_{xxx} . 
\label{xyBL} 
\end{equation} 
The mass conservation (\ref{fluxconservation}) is non-
dimensionalized to 
\begin{equation}  
  h_t + (h v)_x = 0 . 
\label{fluxconsip1}  
\end{equation} 
 
Now, we make an ansatz for the $u$-profile, 
and average over the thickness 
in order to obtain two simplified models. 
First, we use the self-similar velocity profile: 
\begin{equation}  
  u(x,y,t)/v(x,t) = f(\eta) 
\label{ssimprofip}  
\end{equation}   
where $\eta = y/h(x,t)$ and the function $f(\eta)$ satisfies 
\begin{equation}  
 \begin{array}{l} 
   f(0)=0 \\ 
   f'(1)=0 \\ 
   \displaystyle \int_0^1 f(\eta) d\eta = 1 .   
 \end{array} 
\label{fcondip} 
\end{equation}  
Plug this ansatz into (\ref{xyBL}), multiply it by $h$, 
and average over $y$ to obtain 
\begin{equation}  
  (h v)_t + G (h v^2)_x = \frac{3h}{R}  
    - \frac{3}{R \tan \alpha} h h_x - \frac{3v}{Rh} + W h h_{xxx} 
\label{shkadovmod} 
\end{equation}  
together with the mass conservation (\ref{fluxconsip1}). 
Here,  
\[ 
  G = \frac{1}{h} \int_0^h (u/v)^2 dy = \int_0^1 f^2(\eta) d\eta 
\] 
is a constant for a given profile in this model. 
We shall use $G=6/5$ for concreteness, 
corresponding to the parabolic profile $f = 3 (\eta - \eta^2/2)$. 
Equation (\ref{shkadovmod}) is the Cartesian analogue 
of the Kurihara-Tani equation (\ref{Tanieqn}), 
with time-dependent and surface tension terms. 
 
Next, we assume a variable one-parameter profile for $u$. 
As before, we use a third-order polynomial 
\begin{equation} 
  u(x,y,t) = v(x,t) (a \eta+ b \eta^2+ c\eta^3) 
\label{ansatz} 
\end{equation} 
with 
$a = \lambda + 3$, 
$b = -(5 \lambda +3)/2$, 
and $c = 4 \lambda/3$ 
chosen to satisfy the conditions (\ref{fcondip}) for $f$. 
The shape parameter $\lambda(x,t)$ is the single variable 
 characterizing the velocity  profile. 
To describe the evolution of $\lambda(x,t)$ and $h(x,t)$ 
we choose the same set of equations as in the circular hydraulic 
jump. 
The first equation is the mass flux equation (\ref{fluxconsip1}). 
In addition, we use the momentum equation (\ref{xyBL}) 
multiplied by $h$ and averaged in $y$, 
and also (\ref{xyBL}) evaluated at $y=0$: 
\begin{equation} 
 \begin{array}{l} 
  \displaystyle 
  (h v)_t + (h v^2 G(\lambda))_x =  
    \frac{3h}{R} - \frac{3}{R \tan \alpha} h h_x 
    - \frac{v}{Rh} (\lambda+3) + W h h_{xxx} \\   
  \displaystyle 
  0 = \frac{3}{R} - \frac{3}{R \tan \alpha} h_x 
    - \frac{v}{R h^2} (5 \lambda+3) + W h_{xxx} 
 \end{array}  
\label{1parammodel} 
\end{equation} 
where $G(\lambda)$ is given by (\ref{defF2}) as before. 
This system can be cast into the more compact form: 
\begin{equation} 
 \begin{array}{l} 
  \displaystyle 
  (h v)_t + (h v^2 G(\lambda))_x = \frac{4 v \lambda}{Rh} \\   
  \displaystyle 
  h_x \cot \alpha = 1 - \frac{v}{3 h^2} (5 \lambda+3) + 
    \frac{WR}{3} h_{xxx} . 
\end{array}  
\label{shinyaipst}  
\end{equation}  
In the following we call (\ref{shkadovmod}) with 
(\ref{fluxconsip1}) 
the ``similarity model''\footnote{ 
The similarity model is 
 the  ``Shkadov model" considered in \cite{Chang1,Chang2} when 
$W \neq 0$.} 
and (\ref{shinyaipst}) with (\ref{fluxconsip1}) 
the ``one-parameter model''. 
Both models inherit the trivial uniform solution 
from the complete Navier-Stokes model: 
$h = v = q \equiv 1$, and $\lambda \equiv 0$ 
(parabolic profile) for the one-parameter model. 
 
\subsection{Stationary solutions in a moving coordinate frame} 
 
Here, we are concerned with either stationary solutions 
or traveling waves whose surface profiles may show abrupt 
changes. 
Both types of solutions can be sought as stationary solutions 
in a moving coordinate system with a suitable constant velocity $c$, 
including the possibility $c=0$. 
Thus, we use the traveling wave coordinate $\xi = x - ct$, 
and rewrite the models within this frame. 
 
Using the chain rule, 
the mass conservation (\ref{fluxconsip1}) 
used in both models becomes 
\[ 
  - c h_{\xi} + ( h v )_{\xi} = 0 
\] 
which can be integrated to  
\begin{equation}  
  - c h + hv \equiv Q \mbox{(const.)} 
\label{fluxcons1d}  
\end{equation} 
where $Q$ is the mass flux, viewed in the moving 
frame.\footnote{
Note that the flux $q(x,t)$ in the laboratory frame is, in general,
not a constant.
The discharge at the inlet, e.g., at $x=-\infty$
must be varied in time accordingly.
} 
The flow must approach the uniform equilibrium flow $h=1$ 
in the $\xi \rightarrow \infty$ limit. 
Suppose it also approaches another equilibrium flow 
$h=h_2$ in the $\xi \rightarrow -\infty$ limit. 
Then, using (\ref{ndfluxpar}), the condition becomes 
\begin{equation} 
 - c h_2 + h_2^3 = Q = - c + 1 . 
\label{massfluxcondition} 
\end{equation} 
Of course, $h_2=1$ is a solution of this equation. 
In this case we might still be able to find a non-trivial solution 
of a pulse-like solitary wave form. 
Such solutions have previously been studied well 
\cite{Chang1,Chang2}, 
and we do not further seek this type of solutions. 
For a solution of (\ref{massfluxcondition}) other than 
$h_2=1$, we need 
\begin{equation}  
  c = h_2^2 + h_2 + 1. 
\label{consfluxresc} 
\end{equation} 
The solution that can be positive is 
\[ 
  h_2 = \frac{-1 + \sqrt{4 c - 3}}{2} 
\] 
which  is positive if and only if $c>1$. 
 
When $c>1$ two different equilibrium solutions exist, 
and we hope to find a kink-like solution 
which connects the two limiting flows. 
However $c>1$ is only the {\em necessary} condition for  
its existence. 
Sufficiency for the existence depends on the models 
and the parameters: $R$, $\alpha$, and $c$. 
In Sec.~3.5 we shall clarify the parameter regime 
for finding such solutions. 
It turns out that the velocity profiles in this type of 
solutions do not deviate much from parabolic 
even in the one-parameter model. 
In this sense 
they correspond to somewhat ``mild'' jumps 
in terms of the flow structure. 
 
In Sec.~3.6 we find another family of solutions 
which approaches $h=1$ as $\xi \rightarrow \infty$ 
when $c<1$. 
These solutions do {\em not} start from an equilibrium state 
at $\xi=-\infty$. 
Instead, they are only valid for $\xi$ larger than some value  
$\xi_0$. 
In the similarity model they are not interesting  
since they  approach  $h=1$ smoothly. 
However, within the one-parameter model, 
an abrupt change is developed 
in both the surface and velocity profiles, 
sometimes with separation. 
We interpret this solution, when $c=0$,  
as the analogue of the circular hydraulic jump 
in the Cartesian geometry. 
 
The presence of surface tension makes the order of 
the equations higher and makes it more difficult to compute 
the solutions even when they exist. 
We assume that $W$ is small and negligible, 
and set $W=0$ in this section. 
Under this assumption 
we convert the averaged models 
into the moving coordinate frame at velocity $c$. 
Equation (\ref{shkadovmod}) in the similarity model becomes: 
\begin{equation} 
  - c (h v)_\xi + \frac{6}{5} ( h v^2 )_\xi + 
  \frac{3}{R \tan \alpha} h h_\xi = 
  - \frac{3v}{Rh} + \frac{3h}{R} . 
\label{nonstatshwatsol} 
\end{equation}  
Using the condition (\ref{fluxcons1d}), 
$v$ can be eliminated. 
We obtain a first order differential equation for $h$: 
\begin{equation}  
  \frac{dh}{d\xi} = \frac{15}{R}  
  \frac{(h-1)(h^2+h+1-c)}{c^2 h^2 - 6 (1-c)^2 + 15 h^3/(R \tan 
\alpha)} . 
\label{dhshkadov} 
\end{equation} 
Similarly, (\ref{shinyaipst}) in the one-parameter model 
is converted to: 
\begin{equation} 
 \begin{array}{l}  
  \displaystyle 
  - c (h v)_\xi + (h v^2 G(\lambda))_\xi = \frac{4 v \lambda}{Rh} 
\\ 
  \displaystyle 
  h_\xi \cot \alpha = 1 - \frac{v}{3 h^2} (5 \lambda+3) 
\end{array}  
\label{shinyaipsol}  
\end{equation}  
to be solved with (\ref{fluxcons1d}). 
One variable, for instance $v$, can be eliminated 
so that the system becomes two-dimensional 
for $h$ and $\lambda$. 
 
In the following sections we treat 
these averaged models as ``dynamical systems'', 
and view $\xi$ as a time-like variable. 
Fixed points of these systems correspond to the uniform, 
equilibrium solutions of the original time-dependent equations. 
Note that stability in terms of  the variable $\xi$ is not equivalent 
to temporal stability of the original time-dependent equations. 
 
\subsection{Traveling wave solutions} 
\label{sec:travelingwave} 
 
Due to the relationship (\ref{consfluxresc}) 
which is a  one-to-one map between $c$ and $h_2$ in the range $c>1$, 
we may treat $h_2$ or $c$ as the primary parameter 
interchangeably. 
Using $h_2$ as a parameter corresponds physically 
to varying the height and discharge upstream 
and then observing the corresponding change in the wave velocity.  
The condition $c>1$ is equivalent to $h_2>0$, 
and $h_2>1$ if $c>3$. 
The two regimes $h_2>1$ and $h_2<1$ are qualitatively different. 
For $h_2>1$ the discharge at $\xi \rightarrow -\infty$ 
is increased, and a forward-facing front travels downstream. 
As we shall see in this section, 
this state exists for small enough $R$. 
In contrast, $h_2<1$ corresponds to a backward-facing front 
which is found to exist for large enough $R$ but  
seems to us very likely unstable.
Thus, we concentrate on the case  
$h_2>1$ in the following.\footnote{ 
If we used the geometric mean of the up- and downstream heights 
$\sqrt{\tilde{h}_1 \tilde{h}_2}$ as the characteristic length, 
we would obtain equations whose symmetric appearance makes it 
easy to study the forward- and backward-facing fronts 
simultaneously. 
However, we have chosen to scale by the downstream height  
$\tilde{h_1}$ in order to treat the traveling waves 
as well as the stationary jumps. 
} 
 
\subsubsection{The similarity model} 
 
\begin{figure}[t] 
 \centerline{\psfig{file=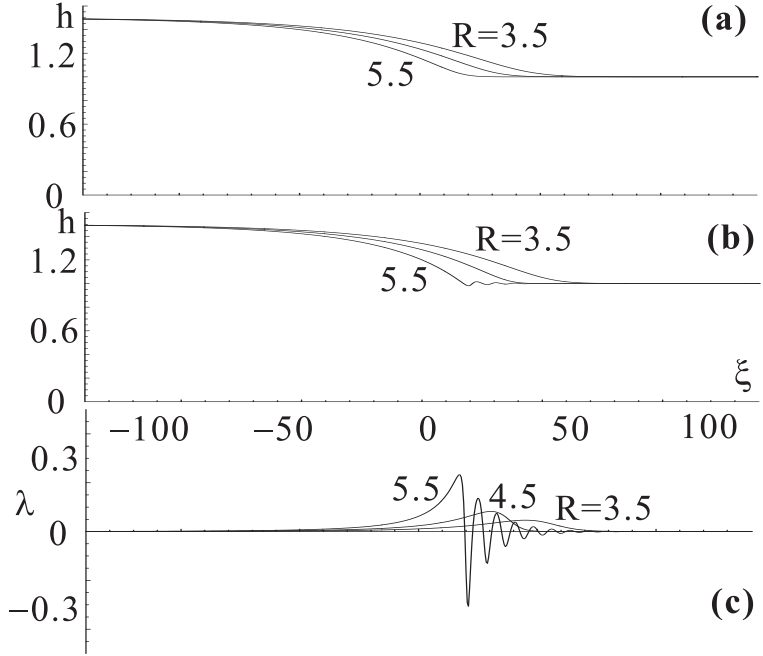,width=10cm}} 
\caption{  
Computed examples of the traveling wave solutions 
connecting two equilibrium states. 
Here, 
the angle of the plane $\alpha=2$[deg], and the height 
$h \rightarrow h_2=1.5$ as $\xi \rightarrow -\infty$, 
corresponding to the front velocity $c=4.75$. 
Three solutions for $R=3.5$, 4.5, and 5.5 are shown. 
(a) 
Height $h$ from  solution 
of the similarity model (\protect{\ref{dhshkadov}}). 
The front becomes steeper as $R$ increases. 
(b) 
Height $h$ from  solution 
of the one-parameter model (\protect{\ref{shinyaipsol}}). 
The curves are quite similar to the ones in (a) 
except for the oscillation in the shallower side 
when $R$ becomes close to a critical value.  (See text.) 
(c) 
Shape parameter $\lambda$ corresponding to the solutions in (b). 
They deviate from the parabolic profile $\lambda=0$ 
and oscillate (for $R=5.5$), but only slightly. 
This explains the similarity between (a) and (b). 
} 
\label{travelingwavegr} 
\end{figure} 
 
Since (\ref{dhshkadov}) is a first order autonomous ordinary 
differential equation, 
the necessary condition for the existence of a heteroclinic orbit 
starting from $h_2 (>1)$ and arriving at $h=1$ 
is that the fixed point $h=1$ is stable and $h_2$ is unstable. 
By linearization, the fixed point $h=1$ is found to be stable if 
\begin{equation} 
  c^2 - 6 (1-c)^2 + 15 / (R \tan \alpha) >0 
\label{stab1shkad0} 
\end{equation} 
or, 
\begin{equation}     
  R \tan \alpha < \frac{15}{6 (1-c)^2 - c^2} = 
  \frac{15}{5 h_2^4+10h_2^3+3 h_2^2- 2 h_2-1} 
  \equiv f_1(h_2) 
\label{stab1shkad}  
\end{equation} 
where the denominator is positive for $c>3$. 
Similarly, $h_2$ is found to be unstable if 
\begin{equation}     
  R \tan \alpha < \frac{15 h_2}{ -h_2^4-2 h_2^3+3 h_2^2 
+10h_2+5} 
  \equiv f_2(h_2) .  
\label{stab2shkad}  
\end{equation}  
The denominator of $f_2$ vanishes only at 
$h_2 = h_2^{\mbox{\scriptsize max}} \approx 2.13$ 
for the region $h_2 > 1$. 
If $h_2 > h_2^{\mbox{\scriptsize max}}$, 
then $f_2 < 0$ and (\ref{stab2shkad}) cannot be satisfied. 
We discard this region of $h_2$. 
For $1 < h_2 < h_2^{\mbox{\scriptsize max}}$ 
one finds that $f_2(h_2) > 1 > f_1(h_2)$. 
Thus, the necessary condition for the existence is 
simply (\ref{stab1shkad}).  
Once the necessary condition is fulfilled, 
sufficiency is guaranteed. 
To see this, 
we only need to ensure that 
the denominator on the right hand side of (\ref{dhshkadov}) 
does not vanish in the region $1 < h < h_2$. 
Suppose it vanished at $h_s$, then we would have 
\begin{equation} 
  c^2 h_s^2 - 6 (1-c)^2 + 15 h_s^3 / (R \tan \alpha) = 0 . 
\label{singularheight} 
\end{equation} 
Comparison with (\ref{stab1shkad0}) gives us 
\[ 
  c^2 (1-h_s^2) + 15 (1-h_s^3) / (R \tan \alpha) > 0 . 
\] 
It is clear that $h_s>1$ is impossible. 
Thus, $h_s<1$, and there is no vanishing denominator in $1 < h < 
h_2$. 
In Fig.~\ref{travelingwavegr}(a)  
we show computed solutions of (\ref{dhshkadov}) 
for three different Reynolds numbers. 
The parameters $\alpha$ and $h_2$ are fixed, 
such that (\ref{stab1shkad}) becomes $R < 6.95$. 
Within this range, a larger $R$ makes the propagating 
front sharper. 
 
\subsubsection{The one-parameter model} 
 
We can eliminate $v$ from 
(\ref{fluxcons1d}) and (\ref{shinyaipsol}), 
and think of trajectories on the 
phase portrait for $(h,\lambda)$. 
We look for a heteroclinic orbit 
starting from a fixed point $(h_2,0)$ 
and arriving at $(1,0)$ 
as $\xi \rightarrow \infty$. 
It is necessary for its existence 
that the point $(h_2,0)$ has 
at least one unstable direction 
and $(1,0)$ has at least one stable direction. 
Linearizing around the  equilibrium point 
as $h = h_e + \delta h$ and $\lambda = 0 + \delta \lambda$, 
where $h_e = 1$ or $h_2$, 
we obtain: 
\[ 
  \left( \begin{array}{c} \delta h_\xi \\ \delta \lambda_\xi 
    \end{array} \right) = J 
  \left( \begin{array}{c} \delta h \\ \delta \lambda \end{array} 
\right) . 
\] 
It is straightforward to calculate 
the $2 \times 2$ Jacobian matrix $J$, 
and show that 
\begin{equation} 
  \det J = \frac{60 (c - 3 h_e^2) \tan \alpha}{R h_e^7} . 
\label{detJ} 
\end{equation} 
For the point $(h_2,0)$ we have 
$c - 3 h_e^2 = 1 + h_2 - 2 h_2^2 < 0$ when $h_2>1$. 
This means that $\det J <0$ for $h_2>1$, 
and the fixed point is always a saddle, 
having exactly one unstable direction. 
 
For the point $(1,0)$ we have $\det J >0$ since 
$c - 3 h_e^2 = h_2^2 + h_2 - 2 > 0$ when $h_2>1$. 
Thus, we must also compute the trace of $J$ for $h_e = 1$ 
which can be shown to be 
\[ 
  \mbox{tr}J = -\frac{60}{R} + ( 33 - 61 c + 25 c^2 ) \tan \alpha . 
\] 
For the stability of $(1,0)$ 
we need tr$J < 0$. 
Since $33 - 61 c + 25 c^2>0$ for $c>3$, 
this condition becomes 
\begin{equation}  
  R \tan \alpha < \frac{60}{33 - 61 c + 25 c^2} = 
    \frac{60}{-3 - 11 h_2 + 14 h_2^2 + 50h_2^3 + 25 h_2^4} 
    \equiv f_s(h_2) . 
\label{exist2}  
\end{equation} 
When this is satisfied, 
the fixed point is locally attracting, 
and a trajectory may reach it from any direction. 
Indeed, we find numerically that 
the condition (\ref{exist2}) also seems to be sufficient. 
For any $R$ and $\alpha$ we have tried in the range (\ref{exist2}), 
a heteroclinic solution was found.  
Computed solutions for three different values of $R$ 
are shown in Fig.~\ref{travelingwavegr}(b) and (c). 
The parameters $\alpha$ and $h_2$ are identical to the ones 
used for the similarity model in Fig.~\ref{travelingwavegr}(a). 
The condition (\ref{exist2}) yields $R<5.59$. 
The height profiles in (b) are essentially identical to the ones in (a). 
This is because the shape parameter $\lambda$ shown in (c) 
does not deviate much from $\lambda=0$, the parabolic profile. 
 
In Fig.~\ref{travelingwavegr}(b) and (c), 
the solution is oscillatory around $h=h_1$ and $\lambda=0$ 
for $R=5.5$. 
This is a feature seen when $R$ becomes 
close to the critical value given by (\ref{exist2}). 
It happens when the type of the fixed point $(1,0)$ 
changes from a stable node to a stable focus. 
The point is a focus when 
$\det J > (\mbox{tr}J)^2/4$, 
which is equivalent to 
$f_+(h_2) < R \tan \alpha < f_-(h_2)$ 
where 
\begin{equation}  
  f_\pm(h_2) = \frac{60}{ -7 - 9 h_2 + 16 h_2^2  
      + 50 h_2^3 + 25 h_2^4 \pm 2 \sqrt{5D}} 
\label{oscillcond} 
\end{equation}  
and 
\[  
  D = 2 + 3 h_2 - 9 h_2^2 - 19 h_2^3 + 3 h_2^4 + 15 h_2^5 + 5 
h_2^6 . 
\] 
It can be seen that 
$f_+(h_2)<f_s(h_2)<f_-(h_2)$ for $h_2>1$. 
Therefore, a heteroclinic solution can be found and 
exhibits oscillations in a small region 
$f_+(h_2) < R \tan \alpha < f_s(h_2)$. 
In Fig.~\ref{travelingwavegr}(b) and (c) 
this condition corresponds to $4.81 < R < 5.59$, 
so only the solution for $R=5.5$ shows oscillations.  
 
\subsection{Stationary jumps} 
 
If $c<1$, the two averaged models have only one fixed point 
$h=1$. 
Therefore, one might imagine that it is too limited to show 
any jump-like structures. 
Nevertheless, we look for trajectories that 
approach to the fixed point as $\xi \rightarrow \infty$. 
Even though  $c=0$ is the physically most 
interesting case, we treat the general case $0 \leq c < 1$. 
Since there is no $h_2$, 
we use $c$ as the prime parameter in this section. 
 
\subsubsection{The similarity model} 
 
The sole fixed point $h=1$ must be stable to be 
the limiting point of a trajectory as $\xi \rightarrow \infty$. 
For $0 \leq c < 1$, the condition is similar to (\ref{stab1shkad0}) 
but with reversed inequality
\begin{equation} 
  c^2 - 6 (1-c)^2 + 15 / (R \tan \alpha) < 0. 
\label{revstab1shkad0} 
\end{equation} 
The singular height $h_s$ of the governing equation 
is still given by (\ref{singularheight}), 
and, using a similar argument as before, 
it is easy to see that $0 \leq h_s < 1$ is impossible when $c<1$. 
Thus, there is a trajectory which approaches $h=1$ from below 
if (\ref{revstab1shkad0}) holds. 
When $1 > c > (6-\sqrt{6})/5 \simeq 0.71$, 
$c^2 - 6 (1-c)^2 >0$ and  
(\ref{revstab1shkad0}) cannot be satisfied. 
When $c < (6-\sqrt{6})/5$, 
the condition is equivalent to 
\begin{equation}  
  R \tan \alpha > \frac{15}{6(1-c)^2-c^2} , 
\label{condhGT1} 
\end{equation}     
which is satisfied in a range of $R \tan \alpha$ 
since the denominator of the right hand side is positive. 
 
\begin{figure} 
 \centerline{\psfig{file=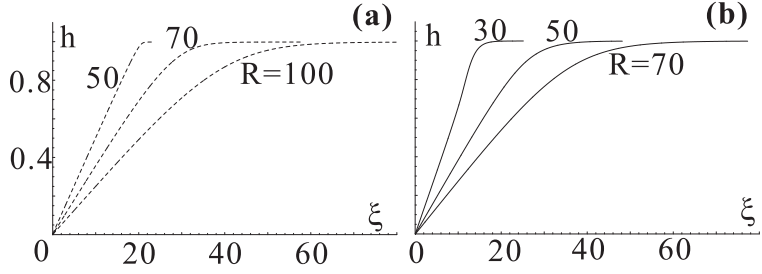,width=10cm}} 
\caption{  
Computed stationary solutions for $\alpha=3$[deg] and $c=0$. 
Dashed curves are solutions 
of the similarity model (\protect{\ref{dhshkadov}}) 
for $R=50$, 70, and 100. 
Solid curves are solutions of the one-parameter 
model (\protect{\ref{shinyaipsol}}) 
for $R=30$, 50, and 70. 
A larger $R$ corresponds to a slower convergence to 
the equilibrium flow $h=1$. 
These solutions do not show any shock-like structure. 
}   
\label{stationaryjumpgr1} 
\end{figure} 
 
Computed solutions for $R=50$, 70, and 100 are shown 
in Fig.~\ref{stationaryjumpgr1} as dashed curves 
using $\alpha=3$[deg] and $c=0$. 
The condition (\ref{revstab1shkad0}) becomes 
$R > 47.7$, and is satisfied for all three. 
Each solution simply approaches $h=1$ smoothly, 
clearly reflecting the first order nature of the model  (\ref{dhshkadov}). 
As $\xi$ decreases, the height vanishes at a finite $\xi$ and 
an inlet must be placed before this happens. 
If $h$ is very small, (\ref{dhshkadov}) simplifies to 
$dh/d\xi = 5/\{2 R (1-c)\}$. 
The solution is 
\begin{equation}  
  h(\xi) = \frac{2.5}{R(1-c)} (\xi-\xi_0) 
\label{hshkadov0} 
\end{equation} 
for some $\xi=\xi_0$ where $h=0$. 
There is no abrupt change in the solutions that 
resembles a stationary shock structure. 
If we use $R$ smaller than the critical value, 
then there is no solution converging to $h=1$. 
Therefore, we view the similarity model as 
inadequate for describing stationary jumps. 
 
\subsubsection{The one-parameter model} 
 
The sole fixed point of this model when $c<1$ 
is $(h,\lambda)=(1,0)$. 
The Jacobian and its determinant  
is still given by (\ref{detJ}), 
but now $c - 3 h_e^2 = c-3 <0$  
and, thus, $\det J <0$. 
Therefore, the fixed point is always a saddle 
in this range of $c$, 
and there is one direction 
convergent to the fixed point as $\xi \rightarrow \infty$. 
It is easy to compute the corresponding trajectory 
by integrating backward in $\xi$ 
from the vicinity of the fixed point. 
This solution seems to exist for all values of $R$, $\alpha$ and 
$c<1$. 
We are interested in solutions which approach $h=1$ from below, 
and  tend to $h=0$ at some $\xi=\xi_0$ as $\xi$ decreases. 
(To be physical, an inlet condition must be specified 
at some $\xi>\xi_0$.) 
We can analyze the solutions asymptotically near $\xi_0$ 
by assuming that $h \sim A (\xi - \xi_0)$ as $\xi \rightarrow \xi_0 
+ 0$. 
Then, using (\ref{fluxcons1d}) and $Q=1-c$ in 
(\ref{massfluxcondition}), 
we obtain $v \sim (1-c)/\{ A (\xi - \xi_0 ) \}$. 
Substituting these into (\ref{shinyaipsol}.2) yields 
\[ 
  \lambda \sim -0.6 + \frac{3 A^3}{5(1-c)} (1-A \cot \alpha) (\xi - 
\xi_0)^3 . 
\] 
Finally, comparing coefficients of the dominant terms 
in (\ref{shinyaipsol}.1) determines $A$ as 
\[ 
  A = \frac{12}{5 R G(-0.6) (1-c)} \approx \frac{1.93}{R (1-c)} . 
\] 
   
\begin{figure} 
 \centerline{\psfig{file=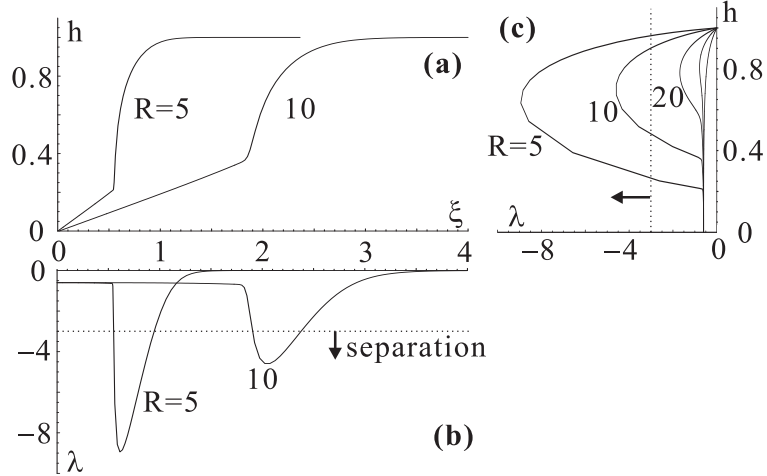,width=10cm}} 
\caption{  
(a) Computed height $h$ of the stationary solutions 
for the one-parameter model (\protect{\ref{shinyaipsol}}) 
using $\alpha=3$[deg], $c=0$, and $R=5$ and 10. 
A shock-like structure is visible, 
with a fast shooting flow in front of it 
and a slow equilibrium flow behind. 
(b) 
The shape parameter $\lambda$ corresponding to the solutions in 
(a) 
shows separation, $\lambda<-3$, in both solutions. 
(c) 
Corresponding trajectories on the phase portrait of $h$ versus 
$\lambda$. 
In addition to the two solutions for $R=5$ and 10, 
three more solutions for $R=20$, 30, and 50 are shown. 
An excursion to small $\lambda$ before convergence to the fixed 
point 
at $(0,1)$ is visible for trajectories with small $R$. 
}   
\label{stationaryjumpgr2} 
\end{figure} 
 
We observe two qualitatively different types 
depending on the parameter values. 
If $\lambda$ increases at the point $\xi=\xi_0$, 
then the solution reaches the parabolic  
profile $\lambda=0$ monotonically. 
This occurs when $R$ is large, 
and three computed solutions are shown in 
Fig.~\ref{stationaryjumpgr1} 
as solid curves. 
The height profile is qualitatively identical to the ones 
from the similarity model shown in dashed curves. 
They do not show any jump structure. 
 
On the other hand, if $\lambda$ decreases at $\xi_0$, 
then the trajectory makes an excursion to smaller $\lambda$, 
sometimes into the separation zone $\lambda<-3$, 
before recovering toward $\lambda=0$. 
The condition to obtain the second type is $A \cot \alpha >1$, or, 
\begin{equation}  
  R \tan \alpha < \frac{12}{5 G(-0.6)}(1-c) \simeq 1.94 (1-c)  
\label{existstatshocks} 
\end{equation} 
with $G(\lambda)$ given by (\ref{defF2}).  
Two solutions satisfying this condition are shown 
in Fig.~\ref{stationaryjumpgr2}(a) and (b). 
Both the height profile and the shape parameter vary 
in a similar manner to the one we obtained 
in the circular hydraulic jump. 
The phase portrait in (c) demonstrates how rapid and large 
the excursion can become for small $R$. 
This type of solution could be realized, for instance, 
as a stationary flow ($c=0$) exiting a sluice gate placed 
at some $\xi > \xi_0$.\footnote{ 
A full-scale channel flow such as a river certainly 
requires a turbulence modelling, 
but we  have been able to construct a miniature experimental model 
in which the flow remains laminar. 
However, our preliminary observation is that 
a pair of edge waves are created from the ends of the gate, 
which makes the flow three-dimensional. 
} 
 
%%%%%%%%%%%%%%%%%%%%%%%%%%%%%%%%%%%%%%%%%%%%%%%%%%%%%%%%%%%%%%%%%%%%%%%%%%%% 
\clearpage 
\section{Linear stability of equilibrium states} 
 
It is quite difficult to carry out 
linear stability analysis around the stationary solutions 
and traveling wave solutions found so far. 
They have non-uniform profiles 
obtained only numerically and  some of the solutions have 
singular points beyond which they cannot be continued. 
Moreover,  
the inlet boundary condition can  strongly affect 
the stability properties of the solutions. 
We shall therefore focus on the linear geometry, 
and only study stability of the equilibrium flow $h \equiv$ const. 
The results are, however, expected to be applicable to 
the equilibrium flow sufficiently far downstream of the jump 
in the stationary solutions 
and to flows sufficiently up- and downstream of the moving front 
in case of the traveling wave solutions. 
Since the dispersion relation scales with the chosen 
characteristic length, as described in Sec.~4.4, 
we only need to consider the flow $h \equiv 1$. 
Both the similarity model (\ref{shkadovmod}) 
and the one-parameter model (\ref{shinyaipst}) are considered, 
including the surface tension term 
which is expected to be relevant \cite{Pumir} for  stability.
One of our aims is, of course, to judge when infinitesimal 
disturbances grow and whey they decay, 
but their propagation velocities are also of our great interest. 
By comparing the velocities to a reference velocity, 
which is zero for the stationary jump and 
$c (>3)$ for the traveling wave, 
we are able to classify different parts of the solutions 
as either super- or subcritical. 
 
\subsection{Dispersion relations} 
 
The first step is to linearize the models 
around the fixed point $h=v=1$ and, 
for the one-parameter model, $\lambda=0$. 
We assume infinitesimal disturbances $\delta h$, $\delta v$, 
and $\delta \lambda$, 
and decompose them into Fourier modes: 
\begin{equation} 
  \delta h, \delta v, \delta \lambda \sim e^{i (k x - \omega t)}.   
\label{ansatzwaves} 
\end{equation} 
 
Plugging them into the linearized equations for the similarity 
model 
(\ref{fluxconsip1}) and (\ref{shkadovmod}), we obtain:  
\begin{equation}  
  \omega^2 +  \omega \left( \frac{3i}{R} - \frac{12}{5} k \right) +  
  \left( -\frac{9i}{R} k + \frac{6}{5} k^2 - \frac{3}{R \tan \alpha} 
k^2  
    - W k^4 \right) = 0 . 
\label{disprelimpl}  
\end{equation} 
Solving the equation the dispersion relation is found to be 
\begin{equation}  
  \omega_\pm = - \frac{3i}{2R} + \frac{6}{5} k \pm \sqrt{D_0} 
\label{exactdisp1}    
\end{equation} 
where the discriminant is 
\begin{equation} 
 D_0 = -\frac{9}{4R^2} +\frac{27i}{5R} k + 3 k^2 \left( 
          \frac{2}{25} + \frac{1}{R \tan \alpha} \right) + W k^4 . 
\label{discrsimilarity} 
\end{equation} 
 
Similarly, from the one-parameter model 
(\ref{fluxconsip1}) and (\ref{shinyaipst}), 
we obtain the dispersion relation: 
\begin{equation}  
  \omega_\pm = - \frac{6i}{5R} + \frac{61}{50} k 
    \pm \frac{3}{5} \sqrt{D_1} 
\label{expldisp}  
\end{equation}  
where 
\begin{equation}  
  D_1 = -\frac{4}{R^2} +\frac{178i}{15 R} k + \left( 
    \frac{421}{900} +\frac{20}{3R \tan \alpha} \right) k^2 + 
    \frac{i}{9 \tan \alpha} k^3 + \frac{20W}{9} k^4 + 
\frac{iRW}{27} k^5 . 
\label{discrexpl}  
\end{equation} 
Note that this model also has only two dispersion relations, 
$\omega_+(k)$ and $\omega_-(k)$ 
because the second equation of (\ref{shinyaipst}) 
does not include time-derivatives. 
 
\subsection{Long wave limit}  
 
We first study the long wave limit $k \rightarrow 0$ 
by taking only the lowest order terms in $k$. 
For the similarity model, 
the dispersion relation (\ref{exactdisp1}) becomes 
\begin{equation}  
  \begin{array}{l} 
    \displaystyle 
    \omega_+ = 3 k + i k^2 ( R - \cot \alpha ) + O(k^3) \\ 
    \displaystyle 
    \omega_-= - \frac{3i}{R} - \frac{3}{5} k  
        - i k^2 ( R - \cot \alpha ) + O(k^3) . 
  \end{array}  
\label{longwave1}  
\end{equation} 
As $k \rightarrow 0$, the group velocities 
$d \omega_+ /d k \rightarrow 3$ and $d \omega_- /d k \rightarrow 
(-3/5)$. 
Therefore, waves corresponding to $\omega_-$ propagate 
upstream, 
and {\em the flow is subcritical irrespective of $R$}. 
By studying the dominant imaginary components of 
$\omega_{\pm}$, 
we also find that the reverse propagating branch $\omega_-$ is 
always stable, 
i.e.\ the disturbances decay, for small enough $k$ 
whereas the forward propagating branch $\omega_+$ is stable 
only for small enough Reynolds number satisfying 
\begin{equation} 
  R \tan \alpha <1 .   
\label{stablong1}  
\end{equation}  
 
The limiting dispersion is identical in the one-parameter model 
apart from numerical coefficients. 
For small $k$,  (\ref{expldisp}) becomes 
\begin{equation}  
  \begin{array}{l}  
    \displaystyle 
    \omega_+ = 3 k + i k^2 ( \frac{5}{4} R - \cot \alpha ) + O(k^3) 
\\ 
    \displaystyle 
    \omega_- = - \frac{12i}{5R} - \frac{14}{25} k 
      - i k^2 ( \frac{5}{4} R - \cot \alpha ) + O(k^3) . 
  \end{array}  
  \label{longwaveshinya}  
\end{equation} 
Thus, the flow is always subcritical 
since the long waves in the $\omega_-$ branch 
propagate upstream with velocity $-14/25$. 
Again, this branch is stable for any $R$ 
while the $\omega_+$ branch is stable only for small Reynolds 
numbers: 
\begin{equation}  
  R \tan \alpha < 4/5 .  
\label{stablong2}  
\end{equation} 
 
\subsection{Intermediate range of $k$} 
 
It is quite unexpected 
that the flow is subcritical for any $R$. 
One would intuitively expect that disturbances cannot propagate 
upstream 
for sufficiently rapid flows. 
An explanation can be made by a more careful study of 
the dispersion relations (\ref{exactdisp1}) and (\ref{expldisp}), 
or, in particular, the discriminants $D_0$ and $D_1$. 
 
We first consider the similarity model. 
If the $O(k^2)$ term dominates in $D_0$, 
then the corresponding group velocities become 
\begin{equation}  
  c_{\pm} = \frac{d \omega_{\pm}}{dk} \approx 
    \frac{6}{5} \pm \sqrt{\frac{6}{25} +\frac{3}{R \tan \alpha}}. 
\label{groupvelinterm} 
\end{equation} 
Both $c_+$ and $c_-$ become positive for 
\begin{equation}  
  R \tan \alpha > 5/2 . 
\label{supercrit1}  
\end{equation}   
We attempt to estimate such a range of $k$. 
For brevity 
we assume $R \tan \alpha \ll 25/2$  
so that the coefficient of $k^2$ in $D_0$ 
can be approximated by $3/(R \tan \alpha)$. 
If the magnitude of the $O(k^2)$ dominates in $D_0$, 
then we must have 
\[ 
  \frac{3k^2}{R \tan \alpha} \gg \frac{9}{4R^2}, ~~ 
\frac{27k}{5R}, ~~ W k^4 , 
\] 
that is, 
\begin{equation} 
  \max \left[ 
    \sqrt{\frac{3 \tan \alpha}{4R}} , \frac{9 \tan \alpha}{5} \right] 
  \ll k \ll \sqrt{\frac{3}{RW \tan \alpha}} . 
\label{krangesimilarity} 
\end{equation} 
Using $R=30$, $\alpha = 5$[deg], and $W=0.01$, for instance, 
the condition (\ref{supercrit1}) and (\ref{krangesimilarity}) 
gives a window $0.16 \ll k \ll 10.7$ 
in which we can hope that the $O(k^2)$ term dominates. 
 
\begin{figure}[t] 
 \centerline{\psfig{file=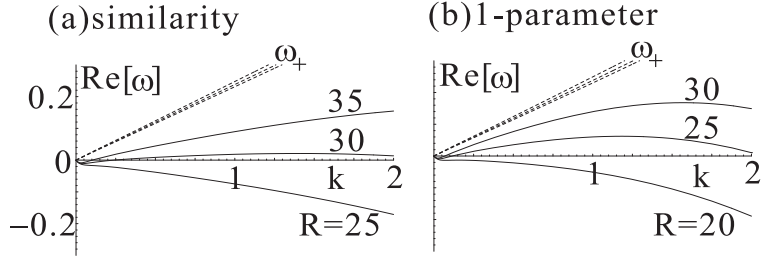,width=10cm}} 
\caption{ 
Real part of the dispersion relation showing the propagation 
of disturbances on the equilibrium flow. 
(a) Similarity model using (\protect{\ref{exactdisp1}}) 
for $R=25$, 30, and 35. 
(b) One-parameter model using (\protect{\ref{expldisp}}) 
for $R=20$, 25, and 30. 
In both models $\alpha=5$[deg] and $W=0.01$ are fixed. 
Three dashed and solid curves correspond to the $\omega_+$ 
and $\omega_-$ branches, respectively, 
of the dispersion relation. 
The $\omega_+$ has a positive slope, 
or group velocity, for all $k$, 
while the $\omega_-$ branch has positive slope 
only when $R$ is large. 
However, for large enough $R$, 
the region of $k$ in which both branches have positive slopes 
extends  from small $k$ corresponding to wavelengths 
beyond the system size to  large $k$ with wavelengths 
smaller than the thickness of the flow. 
In this case the flow is essentially supercritical 
since disturbances are all carried away downstream. 
} 
\label{dispaver_real} 
\end{figure} 
 
Rather than attempting a more accurate estimate of the zone, 
we demonstrate that such an interval can be in fact quite long, 
by plotting the real part of $\omega_{\pm}(k)$ for 
(\ref{exactdisp1}) 
in Fig.~\ref{dispaver_real}(a). 
Three different values of $R$ are used 
while $\alpha$ and $W$ are fixed. 
The $\omega_+$ branch, shown as dashed curves, 
has a positive slope for any $k$. 
Both phase and group velocities of this branch are positive. 
On the other hand, the $\omega_-$ branch, shown as solid curves, 
qualitatively changes with $R$. 
For $R=25$ its slope appears to be negative for all $k$, 
indicating a subcritical flow. 
However, for a larger $R$ there is an interval of $k$ 
in which the slope becomes positive. 
In the limit $k \rightarrow 0$, 
the branch still has a negative slope 
in accordance with the analysis of the long wave limit 
in the previous section. 
However, the subcritical region near $k=0$ can be very small. 
One sees in Fig.~\ref{dispaver_real}(a) that 
the curve has a positive slope already when $k > 0.05$ and 
$R=35$. 
The slope continues to be positive until $k=2$, 
corresponding to a wavelength of half the thickness 
of the equilibrium flow. 
Since the system length is finite in practice, 
the subcritical flow in the $k \rightarrow 0$ limit 
{\em cannot} be achieved, 
and the flow becomes essentially supercritical 
for all the wave numbers observed. 
This defines the super- and subcritical flows 
within our viscous model, 
and confirms the intuitive picture of having a supercritical flow 
when the flow is sufficiently rapid. 
 
The situation is qualitatively identical 
in the one-parameter model. 
We obtain 
\begin{equation}  
  R \tan \alpha > 20/11 
\label{supercrit2} 
\end{equation}   
and 
\begin{equation} 
  \max \left[ 
    \sqrt{\frac{3 \tan \alpha}{5R}} , \frac{50 \tan \alpha}{89} 
\right] 
  \ll k \ll \min \left[ 
    \frac{60 \tan \alpha}{R} , \sqrt{\frac{3}{RW \tan \alpha}} , 
    \left( \frac{180}{R^2 W \tan \alpha} \right) \right] 
\label{krange1param} 
\end{equation} 
as the corresponding equations to (\ref{supercrit1}) 
and (\ref{krangesimilarity}), respectively. 
Again using $R=30$, $\alpha = 5$[deg], and $W=0.01$, 
the interval becomes $0.05 \ll k \ll 0.18$. 
The upper limit comes from the $O(k^3)$ term in $D_1$, 
and is estimated to be rather small 
since we have only compared the magnitudes. 
In fact, when we plot the real part 
of the dispersion relation (\ref{expldisp}) 
in Fig.~\ref{dispaver_real}(b), 
we find that the $\omega_-$ branch has a positive group velocity 
for a much longer range of $k$. 
The supercritical flow near the $k=0$ limit is very small once 
again 
if $R$ becomes as large as $R=25$. 
 
\subsection{Super- and subcriticality for moving fronts} 
 
The intermediate-$k$ behavior enables us to decide 
whether a given equilibrium flow is 
``inherently'' super- or subcritical. 
This distinction is made based on wave velocities 
with respect to the laboratory frame. 
A more classical distinction of the two types 
arises in the context of the shock theory, 
as reviewed in Sec.~2.1. 
In this case velocities are measured with respect to a moving front; 
we call the flow ``supercritical'' 
if the group velocity of all the waves is less than 
the front velocity $c$, 
and ``subcritical'' if there is a wave component 
whose group velocity is larger than $c$. 
Here, we briefly note that 
the averaged equations can describe this traditional 
classification, too. 
 
Take a moving front such as the one shown in 
Fig.~\ref{travelingwavegr}. 
We concentrate on the long wave limit $k \rightarrow 0$. 
For $\xi \rightarrow \infty$ the flow approaches an equilibrium 
flow 
with $h=1$. 
Linear waves propagate forward and backward  
with the group velocities $d \omega_+ /dk=3$ and $d \omega_-
/dk=-3/5$ 
according to the dispersion relation 
for the similarity model (\ref{longwave1}). 
This is a subcritical situation in the laboratory frame, 
but, since the front velocity is $c = 1+h_2+h_2^2 > 3$, 
both these waves propagate into the front. 
Therefore, the flow is supercritical with respect to the front. 
 
To derive the dispersion relation of the equilibrium flow 
with height $h_2$ for $\xi \rightarrow -\infty$, 
consider rescaling the height by $h_2$. 
That is, we use this height as the characteristic length 
so that a wave number $k$ must be multiplied by $h_2$. 
Since the flow rate is $q_2 = h_2^3$ from (\ref{ndfluxpar}), 
the velocity has to be scaled by $q_2/h_2=h_2^2$. 
Thus, the group velocities for this flow 
in the laboratory frame are  
$d \omega_+ /dk = 3 h_2^2$ and $d \omega_- /dk = -(3/5) h_2^2$. 
It is easy to show that $3 h_2^2 > c = 1+h_2+h_2^2$ for $h_2>1$. 
Thus, one wave component propagates into the front 
while the other moves away from it 
so that the flow behind the front is subcritical.  
 
Therefore, the moving front has a supercritical flow 
on the shallower side and a subcritical flow on the deeper side, 
and can be regarded as a classical shock. 
Using the one-parameter model instead of the similarity model 
is qualitatively identical. 
 
\subsection{Short wave limit} 
 
We now come back to the stationary equilibrium flow, 
and study the dispersion relation in the short wave range. 
Since the derivation of the averaged equations relies on 
the assumption of predominantly horizontal flow, 
it is not our aim to accurately resolve 
wave components when $k$ is large. 
We only hope that the short waves decay 
so that they do not  interfere with meaningful dynamics 
when we simulate the time-dependent model. 
Unfortunately, the one-parameter model performs poorly 
in this respect compared to the similarity model. 
 
The dispersion relation of the similarity model (\ref{exactdisp1}) 
can be approximated in the large $k$ limit as 
\begin{equation}  
  \begin{array}{l} 
    \mbox{Re\,} \omega_{\pm} = \pm \sqrt{W} k^2+ O(k) \\ 
    \displaystyle 
    \mbox{Im\,} \omega_{\pm} = -\frac{3}{2 R}+O(1/k).  
  \end{array}  
\label{shortwave1}  
\end{equation}  
Thus, short waves in (\ref{shkadovmod}) are damped out if 
$W>0$.  
 
If we neglect the surface tension and set $W=0$, 
the dispersion relation for large $k$ is 
\begin{equation}   
  \omega_{\pm} = c_{\pm} k - \frac{3i}{2R} \frac{c_{\pm}-
3}{c_{\pm}-6/5}  
  +O(k^{-1}) 
\label{shortwave0}  
\end{equation} 
where $c_\pm$ is the velocity of the corresponding wave given by  
\begin{equation}  
  c_\pm = \frac{6}{5} \pm \sqrt{ \frac{6}{25} +\frac{3}{R \tan 
\alpha} } . 
\label{velshortwave1}  
\end{equation} 
Since $c_-<6/5$ from (\ref{velshortwave1}), 
the branch $\omega_-$ is always stable, 
as can be seen from (\ref{shortwave0}). 
On the other hand, 
since $c_+>6/5$, the condition for the stability of  
the branch $\omega_+$ is $c_+<3$, which is equivalent to  
\begin{equation}  
  R \tan \alpha <1 . 
\label{stabomplus2}  
\end{equation}   
For a large $R$ the equilibrium state is no longer stable, 
but this is reasonable in the absence of surface tension. 
 
Now, we turn into the dispersion relation of the one-parameter 
model 
(\ref{expldisp}). 
For large $k$, it behaves as 
\begin{equation}  
  \omega_\pm \sim \pm k^{5/2} \sqrt{iW/75} \qquad \mbox{if 
$W>0$,} 
\label{largekdisp}  
\end{equation}  
and as 
\begin{equation}  
  \omega_\pm \sim \pm k^{3/2} \sqrt{i/(25 \tan \alpha)} 
  \qquad \mbox{if $W=0$.} 
\label{largekdisp0}  
\end{equation}  
In either case 
one of the branches has an unstable component as $k \rightarrow 
\infty$, 
irrespective of $R$ or $\alpha$. 
We have been unable to find a natural modification to the one-parameter 
model 
which prevents this unphysical behavior. 
Its cause may well be that the evolution of short waves 
is not well represented 
by the boundary layer approximation we started with. 
In fact, in the boundary layer equations (\ref{BLeqinpl}) 
the higher order derivatives of $x$ 
that are thought to be crucial for stability 
of the high-$k$ modes are neglected. 
In this view 
the similarity model (\ref{shkadovmod}) provides 
surprisingly reasonable behaviour for large $k$, 
even starting from (\ref{BLeqinpl}). 
 
%%%%%%%%%%%%%%%%%%%%%%%%%%%%%%%%%%%%%%%%%%%%%%%%%%%%%%%%%%%%%%%%%%%%%%%%%%% 
\clearpage 
\section{Conclusions}  
 
In this article we have presented a simple but fairly quantitative 
method of reducing flows with strongly deformed free
surfaces to a manageable system of equations. 
By assuming a ``flexible'' velocity profile whose shape parameter 
is another dependent variable, 
flows with an internal eddy can be described. 
In the radial geometry our results compare well 
with experiments 
and we have obtained analytic expressions for the circular hydraulic jump. 
 
We have also studied the flow down an inclined plane. 
The reduced equations possess not only the traveling wave 
solutions (heteroclinic orbits) studied  previously 
but also stationary jump solutions. 
We have found that the stationary solutions
show a stronger change in the velocity profile 
than the traveling waves. 
 
Finally, we have classified different parts of the flows 
into super- and subcritical by studying the dispersion relation 
around the equilibrium flow. 
This classification is  standard for inviscid 
shallow water flow and in shock theory, but is is not obvious 
 in the context of viscous flow.  
Indeed, for sufficiently long waves the averaged equations 
show that supercritical flow is not possible. 
However, waves with intermediate lengths can 
make the flow essentially supercritical. 
 
The only but serious defect of our reduced model 
which we have been unable to overcome is 
its short wavelength behavior. 
As it stands now, 
some artificial dissipation term to stabilize 
the short waves is necessary 
before time-dependent simulations are attempted. 
To our dismay  
a more natural treatment of this  problem has so far eluded us. 
 
\section*{Acknowledgements}  
 
The core part of this work was carried out 
while authors were at the Center for Chaos \& Turbulence Studies 
(CATS) 
at the Niels Bohr Institute to which  SW and VP are grateful  
for  hospitality and an inspiring environment.  
SW thanks the Institute for Mathematics \& its Applications (IMA) 
of the University of Minnesota 
for providing him with a place and atmosphere to continue the 
work. 
Research supported in part 
under Grant-in-Aid for Scientific Research of JSPS. VP acknowledges 
the hospitality of University of Chicago and support through the 
NSF grant No. DMR 9415604 and MRSEC, NSF Grant No. DMR 9808595.

\end{document}